\documentclass[
twocolumn
]{aastex62}
\usepackage{graphics} % for pdf, bitmapped graphics files
\usepackage{epsfig} % for postscript graphics files
\usepackage{mathptmx} % assumes new font selection scheme installed
\usepackage{times} % assumes new font selection scheme installed
\usepackage{amsmath} % assumes amsmath package installed
\usepackage{amssymb}  % assumes amsmath package installed
\usepackage{xcolor}
\usepackage{subfigure}
\graphicspath{{./}}

\accepted{December 21, 2021}

\shorttitle{Intermittency in the Expanding Solar Wind: PSP, Helios 1, and Voyager 1}
\shortauthors{Cuesta et al.}
\begin{document}

\title{\bf Intermittency in the Expanding Solar Wind: Observations from Parker Solar Probe (${\bf 0.16~{\rm \bf au}}$), Helios 1 (${\bf 0.3 - 1~{\rm \bf au}}$), and Voyager 1 (${\bf 1 - 10~{\rm \bf au}}$)}

\correspondingauthor{Manuel Enrique Cuesta}
\email{mecuesta@udel.edu}

\author{Manuel Enrique Cuesta}
\affiliation{University of Delaware,
Department of Physics and Astronomy,
Newark, DE 19716, USA}

\author{Tulasi N. Parashar}
\affiliation{Victoria University of Wellington,
Wellington 6410, New Zealand}

\author{Rohit Chhiber}
\affiliation{University of Delaware,
Department of Physics and Astronomy,
Newark, DE 19716, USA}
\affiliation{NASA Goddard Space Flight Center,
Greenbelt, Maryland 20771, USA }

\author{William H. Matthaeus}
\affiliation{University of Delaware,
Department of Physics and Astronomy,
Newark, DE 19716, USA}



\begin{abstract}
    We examine statistics of magnetic field vector components to explore how intermittency evolves from near Sun plasma to radial distances as large as $10~{\rm au}$.  
    Statistics entering the analysis include auto-correlation, magnetic structure functions of order n (SF$_n$), and scale dependent kurtosis (SDK), each grouped in ranges of heliocentric distance.  
    The Goddard Space Flight Center Space Physics Data Facility (SPDF) provides magnetic field measurements for resolutions of $6.8~{\rm ms}$ for Parker Solar Probe, $6~{\rm s}$ for Helios, and $1.92~{\rm s}$ for Voyager 1.  
    We compute SF$_2$ to determine the scales encompassing the inertial range and examine SDK to investigate degree of non-Gaussianity.  
    Auto-correlations are used to resolve correlation scales.  
    Correlation lengths and ion inertial lengths provide an estimate of effective Reynolds number (${\rm R_e}$).  
    Variation in ${\rm R_e}$ allows us to examine for the first time the relationship between SDK and ${\rm R_e}$ in an interplanetary plasma.  
    A conclusion from this observed relationship is that regions with lower ${\rm R_e}$ at a fixed physical scale have on average lower kurtosis, implying less intermittent behavior.  
    Kolmogorov refined similarity hypothesis is applied to magnetic SF$_n$ and kurtosis to calculate intermittency parameters and fractal scaling in the inertial range.
    A refined Voyager 1 magnetic field dataset is generated.\newline
\end{abstract}



%%%%%%%%%%%%%%%%%%%%%%%%%%%%%%%%%%%%%%%%%%
%%%%%%%%%%%%%%%%%%%%%%%%%%%%%%%%%%%%%%%%%%
\section{Introduction} \label{sec:intro}

    There are a variety of motivations for studying solar wind turbulence, ranging from its influence on macroscopic processes such as heating and acceleration of the solar wind, to the cascade and kinetic processes that are responsible for dissipation, and even its influence on energetic particle populations of solar and galactic origin \citep{MatthaeusVelli2011}. 
    During its expansion, the solar wind displays a wide range of plasma and turbulence properties that mutually influence the interaction between the turbulence and the ambient interplanetary environment.
    This offers an opportunity to study how plasma turbulence at magnetohydrodynamic (MHD) scales responds to these varying conditions. 
    The motivation for this project is to explore turbulence intermittency in the solar wind and how it evolves from the inner heliosphere ($\approx 0.16~{\rm au}$) out to $10~{\rm au}$ using in situ measurements.
    Taking advantage of the parameter variations experienced over changing heliocentric distances, in this paper, we investigate the evolution of Reynolds number-related properties in relation to intermittency statistics in the interplanetary magnetic field.
    The study employs measurements by NASA's Parker Solar Probe (PSP), Helios 1, and Voyager 1 spacecraft that collectively probe heliocentric distances that vary from near 20 to about 2000 solar radii.
    These intermittency statistics include structure functions of order n (SF$_n$) and scale dependent kurtosis (SDK), as well as the effective Reynolds number (${\rm R_e}$).

    The results presented will demonstrate that the effective Reynolds number, related to the system size as seen by the turbulence, undergoes a systematic variation from inner to outer heliosphere, with a general statistical trend toward smaller values. 
    The energy-related statistics such as second-order structure functions maintain a form that suggests the turbulence is well-developed at all distances observed in the study.
    The selected intermittency measure that we study, the scale dependent kurtosis, is evaluated using two physically relevant normalizations -- in one case normalized to the outer, energy-containing scale, and in the other, normalized to the inner, or kinetic scale. 
    We find that the kurtosis varies with the Reynolds number, becoming less intermittent in this sense as one moves towards the outer heliosphere. 
    The latter result is qualitatively in accord with well-known behavior of the kurtosis with varying Reynolds number in hydrodynamics. 
    This comparison extends further the general parallelism between turbulence in ordinary fluids and the behavior of turbulence in weakly collisional space plasmas.
    
    Preliminary results involving SDK and the effective Reynolds number ${\rm R_e}$ computed from a refined 
    Voyager 1 magnetic field dataset have been published \citep{ParasharEA2019}.
    Expanding on the prior findings, here we show that ${\rm R_e}$ decreases as the solar wind expands, adopting the outer scale as the correlation scale ($L$) and associating the inner scale to the ion inertial scale (${\rm d_i}$).  
    We also find that the SDK held at $10~{\rm d_i}$ decreases with increasing heliocentric distance $R$ above $1~{\rm au}$, suggesting weaker intermittency at larger $R$ but smaller ${\rm R_e}$. 
    
    The present paper also extends these earlier findings to the inner heliosphere in order to better examine the relationships between ${\rm R_e}$ and SDK. 
    Helios 1 and PSP data are used for this purpose. 
    Prior to presenting the results in Section \ref{sec:results}, all of the relevant quantities and methods are defined in Section \ref{sec:defs} and the datasets employed are described in Appendix \ref{app:data}.
    Discussion and conclusions are given in the final Section \ref{sec:conclusions}.

\section{Background} \label{sec:background}

    Travelling outwards from the solar corona, the solar wind expands to fill in the increasing volume of the heliosphere.
    At the same time the solar wind plasma, behaving in many ways as a turbulent magneto-fluid (although it is not a strongly collisional gas) supports a cascade, with the implication that its larger structures break down into smaller structures, or eddies \citep{Pope2000}.      These complementary effects occur simultaneously to modify the scales of solar wind turbulence; turbulent structures breaking up into smaller structures due to cascade as their relative size increases due to expansion.

    However, in the cascade the production of small scale magnetic field fluctuations due to local nonlinear interactions does not occur uniformly. 
    Magnetic field fluctuations and independently evolving eddies, upon their mutual encounter and interaction, can produce coherent structures such as sheets of electric current, current cores, vortices and density structures \citep{GoldsteinEA1995, Pope2000, AlexandrovaEA2008}. Small scale structure can form, due to, for example, a collision of flux tubes leading to reconnection. 
    This type of process leads to intermittency of turbulence, a term that qualitatively refers to the irregular alternation between regions of strong increments (or gradients) and regions of weaker spatial increments.
    More formally, intermittency refers to the presence of non-Gaussian statistics of certain quantities such as increments of magnetic field \citep{Frisch1995}.

    Such irregular variations are known to occur in MHD fluids and by analogy can occur in the solar wind.  
    However, there is an intrinsic competition between the expansion and the evolution of the turbulence, aforementioned, that can affect expectations regarding intermittent behavior.
    Several effects accompany solar wind outflow and expansion. 
    The turbulence amplitude itself can evolve, increasing near the Alfv\'en critical point \citep{ChhiberEA2019March,AdhikariEAOct2020} and subsequently decreasing, on average, except where driven strongly by shear or other effects. 
    Constant speed expansion also causes transverse outer length scales to increase, and density to decrease. 
    Meanwhile the turbulence ``ages'' in terms of eddy turnover times \citep{MatthaeusEAPRL1998}, and may or may not be immediately or always in a ``fully developed'' state \citep{Pope2000}. 
    The competition involving these effects is complex and it is not obvious how to assess the radial development of physical properties such as intermittency. 
    We address that issue here, with a concentration on development of the correlation scale, the effective Reynolds number, and at a given physical scale, the energy content and the kurtosis, the latter being a measure of intermittency. 
    We note in passing that previous studies have carried out surveys of turbulence properties using Helios and Voyager datasets (e.g., \cite{ZankEA1996,SmithEA2001,GrecoEA2012,PineEA2020b}, and these represent antecedents to the present work, even if the goals and techniques were somewhat different. 

\section{Physical Quantities and Analytic Methods}\label{sec:defs}

    We define the magnetic field fluctuations, $\Vec{b}$, to be the difference of the total magnetic field and the averaged magnetic field $\Vec{B}$ $- < \Vec{B} >$, where $<...>$ represents a suitable averaging technique.  
    The increment of $\Vec{b}$ at a time lag $\tau$ is then $\Delta \Vec{b}(t,\tau) = \Vec{b}(t+\tau) - \Vec{b}(t)$. 
    We will sometimes refer to this simply as 
    $\Delta \Vec{b}$ when it is convenient and does not cause confusion. 
    For all analyses shown here, $\tau$ is implicitly associated with a spatial lag $\ell$ at mean solar wind speed $V_{\rm SW}$ according to $\ell = - V_{\rm SW}\tau$ \citep{Taylor1938}. 
    For data specifics, see Appendix \ref{app:data}.

\subsection{Ion Inertial Length}\label{subsec:di}

    The ion inertial length ${\rm d_i}$ is defined as:
    
    \begin{equation}\label{eq:di}
        {\rm d_i} = \frac{c}{\omega_{pi}} = \frac{228}{\sqrt{n_i}} [{\rm km}]
    \end{equation}
    where $c$ is the speed of light, $\omega_{pi}$ is the ion plasma frequency, and $n_i$ is the proton density in units of ${\rm cm}^{-3}$.  
    Since $n_i(R)$ is known to scale, on average, as $R^{-2}$ in the solar wind, then ${\rm d_i}$ should scale, on average, as $R$.
    Variation in ${\rm d_i}$ is expected due to a variety of factors, including solar cycle, stream structure, and solar events such as coronal mass ejections, all of which cause variability in the proton density.
    These sources of variability will also affect other plasma properties including the magnetic field and proton speed.

\subsection{Auto-Correlation Function and Correlation Length}\label{subsec:corr}

\begin{figure}[htp!]
        \centering
        \includegraphics[scale=.5]{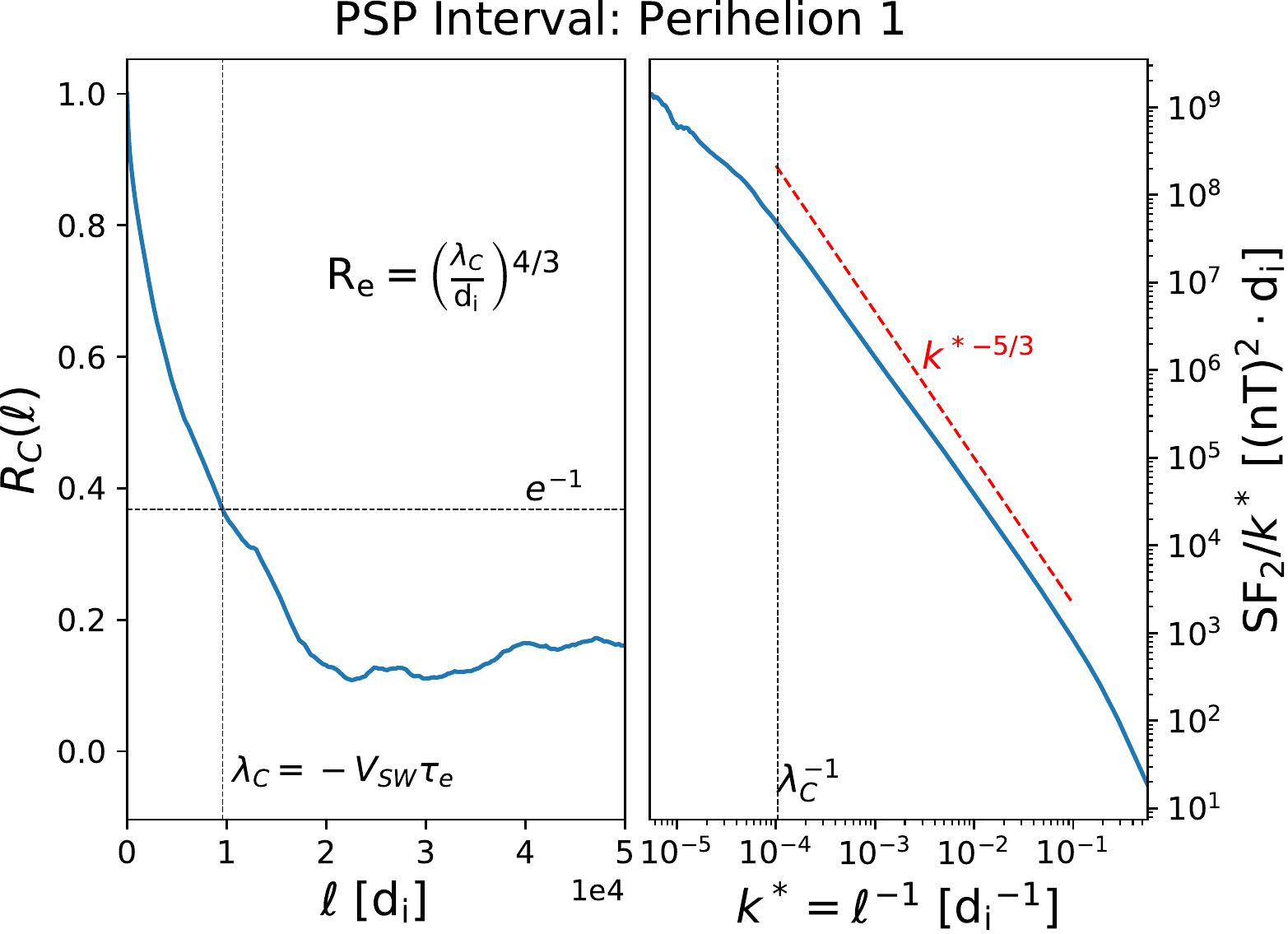}
        \caption{Auto-correlation and equivalent magnetic energy spectrum for PSP's first perihelion.  On the left panel we have $R_C$ as a function of spatial lag $\ell$, in units of ${\rm d_i}$, by applying the Taylor hypothesis to convert temporal lags $\tau$ to spatial lags $\ell$.  The intersection of the horizontal dashed line and $R_C$ indicates the "e-folding" correlation scale.  On the right panel we have the equivalent magnetic energy spectrum in units of $[(nT)^2 \cdot {\rm d_i}]$ as a function of wave-number $k^*=\ell^{-1}$ in units of ${\rm d_i}^{-1}$.  The vertical line in both panels shows $\lambda_C^{-1}$.  Expressed in left panel is the effective Reynolds number definition, ${\rm R_e}=\left( \frac{\lambda_C}{\rm d_i} \right)^{4/3}$, which is further discussed in Section \ref{subsec:Re}.}
        \label{fig:corr}
    \end{figure}

    A quantity of central interest is the two-time correlation function of the magnetic field fluctuations.
    The two-time correlation $R_C$ of time stationary magnetic field fluctuations is defined as:
    
    \begin{equation}\label{eq:corr}
        R_C(\tau) = \frac{\left< \Vec{b}(t) {\bf \cdot} \Vec{b}(t+\tau) \right>}{\left< \Vec{b}(t) {\bf \cdot} \Vec{b}(t) \right>}
    \end{equation}
    where again $\left< ... \right>$ represents a suitable averaging technique over $t$.
    Figure \ref{fig:corr} shows the auto-correlation and equivalent magnetic energy spectrum for an interval near PSP's first perihelion.
    
    In the left panel of Figure \ref{fig:corr} we show how to get the spatial correlation scale.  
    The correlation time $\tau_e$ is a measure of the characteristic time separation over which the magnetic fluctuations become uncorrelated. 
    Here we identify $\tau_e$ with the ``e-folding'' time, i.e., $R_C(\tau = \tau_e) = 1/e$.  
    This correlation time corresponds to the size of the energy containing eddies, within the interval used for computation.
    However, to extract a length $\lambda_C$ from this correlation time scale, a conversion is required.  
    Taylor's hypothesis, $\ell = - V_{\rm SW}\tau$, can be used to convert time lags $\tau$ to spatial lags $\ell$ \citep{Taylor1938}.  
    As a result, the correlation length is defined to be $\lambda_C = - V_{\rm SW}\tau_e$. 
    This approximate conversion of time lags to spatial lags is expected to be accurate when $V_{\rm SW}$, the bulk flow speed of the plasma, is large compared to the characteristic speeds of the local fluctuation dynamics, such as the turbulence speed or the Alfv\'en speed \citep{Jokipii1973}.  
    Applications of the Taylor hypothesis has been used for a variety of statistics such as means and correlations, working well down to scales smaller than the ion inertial scale \citep{MatthaeusEA1982,AlexandrovaEA2008,BrunoEA2013,ChhiberEA2018}.

    A general expectation that $\lambda_C$ scales radially as a power-law may be motivated by appeal to von Karman similarity theory \citep{KarmanHowarth1938} in an inhomogeneous radially expanding medium. 
    As summarized in \citet{ParasharEA2019}, the relationship between the turbulence amplitude $Z$, and the similarity scale $L$ \citep{ZankEA1996,BreechEA2008}
    may be described by the coupled ordinary differential equations $\frac{dZ^2}{dt}=\frac{-Z^3}{L}$ and $\frac{dL}{dt}=\frac{Z}{2}$ \citep{MatthaeusEA1996}, ignoring expansion.
    Note that we consider interval sizes that are small compared to the local heliocentric distance over the range covered in our analysis.
    Therefore, expansion does not strongly affect the observations for any given interval.
    As a result, one can find that $L(t) \sim t^{1/2}$.
    The use of Taylor's hypothesis along with the association of the correlation scale with the similarity scale leads to an expectation that $\lambda_C(R) \sim R^{1/2}$.
    The dependence of the correlation scale on heliocentric distance $R$ has been observed to behave as $\lambda_C(R) \sim R^{0.44}$ in a mixed latitude ensemble, with no strong dependence on plasma beta \citep{KleinEA1992,ZankEA1996,RuizEA2014}.
    
    In Appendix \ref{app:data}
    we will discuss additional features of the 
    data analysis procedures that provide accurate
    estimates of the correlation scale.
    These estimates will then be used to compute ${\rm R_e}$, defined in Equation \eqref{eq:Re}.
    From the equivalent magnetic energy spectrum (EMES; see, e.g., \citet{ChhiberEA2018}) in the right panel of Figure \ref{fig:corr}, the inertial range is approximated to be between $\lambda_C$ and ${\rm d_i}$, for which EMES behaves nearly as a power-law (${k^*}^{-5/3}$).
    Larger values of ${\rm R_e}$ are the result of larger separation between $\lambda_C$ and ${\rm d_i}$, i.e. larger inertial ranges (e.g., \citet{Pope2000}).

%%%%%%%%%%%%%%%%%%%%%%%%%%%%%%%%%%%%%%%%%%
\subsection{Effective Reynolds Number}\label{subsec:Re}

    The correlation and ion inertial scales are utilized in estimating the effective Reynolds Number ${\rm R_e}$, which is a measure of system size from a turbulence perspective.  
    Below, the behavior of ${\rm R_e}$ will be 
    of interest in examining the 
    macroscopic variations of intermittency measures within the heliosphere as measured by Voyager.
    The hydrodynamic value of the Reynolds number, $R_H$, is equal to $uL/\nu$, where  $u$ is the turbulence speed, $L$ is the correlation scale mentioned in Section \ref{subsec:corr}, and $\nu$ is the kinematic viscosity ordinarily well determined for a collisional medium.
    However, for the solar wind within the inner $\approx 10~{\rm au}$ of the heliosphere, $\nu$ is essentially
    meaningless as the plasma is weakly collisional \citep{VerscharenEA2019}.
    
    However, another view of the 
    Reynolds number in Kolmogorov theory is 
    that it quantifies 
    the ratio of the outer scale to an inner scale,
    through a relation that we adapt to 
    define the effective Reynolds number, 
    namely \citep{Pope2000}:
    
    \begin{equation} \label{eq:Re}
        {\rm R_e} = \left(\frac{\lambda_C}{\rm d_i}\right)^{4/3}.
    \end{equation}
    Note that here we 
    approximate the inner scale by 
    the ion inertial scale for a weakly collisional plasma.
    A visualization of this ratio can be found on the right panel of Figure \ref{fig:corr}.
    The value of ${\rm R_e}$ directly depends on the spatial separation between $\lambda_C$ and ${\rm d_i}$.
    Using the expectations
    developed above for the radial behavior of 
    the correlation length and the ion inertial length, 
    we then expect that ${\rm R_e}$ should scale as $R^{-2/3}$.
    This means that the scale separation between the correlation length and ion inertial length is expected to decrease with increasing heliocentric distance.

%%%%%%%%%%%%%%%%%%%%%%%%%%%%%%%%%%%%%%%%%%
\subsection{Structure Function}\label{subsec:SF}

    The (two point) structure functions contain a wealth of information about the structure and intensity of a turbulent field \citep{Pope2000} and are usually written as a function of spatial lag $\ell = -V_{SW} \tau$. 
    The Kolmogorov Refined Similarity Hypothesis (KRSH) \citep{Kolmogorov1962,Obukhov1962} written for longitudinal increments of the velocity, $\Delta v_\ell = \vec{\ell} \cdot \Delta \vec{v}$, postulates that
    
    \begin{equation}
        \Delta v_\ell \sim \epsilon_\ell^{1/3} \ell^{1/3}
        \label{eq:KRSH},
    \end{equation}
    where $\epsilon_\ell$ is the dissipation rate averaged over a volume of order $\ell^3$.
    This relation may be provisionally adapted to magnetic field increments as a surrogate for Els\"asser increments \citep{PolitanoEA1998,WanEA2012,ChhiberEA2021April}. 
    Here we will analyze the structure functions of the magnitude of magnetic vector increments in place of the longitudinal velocity increments entertained in Kolmogorov's hypothesis. 
    We should note that the KRSH for MHD (or plasma) has not been fully demonstrated or confirmed as far as we are aware (see, however, \citet{MerrifieldEA2005, ChandranEA2015}). 

    The set of magnetic field structure functions of order $n$, for varying increment scales $\tau$, may then be defined as:
    
    \begin{equation} \label{eq:sf}
        {\rm SF_n}(\tau) = \left< | \Delta \Vec{b}(t,\tau) |^n \right>.
    \end{equation}
    For the purpose of further analysis below we shall assume that the moments of the magnetic increments, under the modified KRSH, behave as 
    
    \begin{equation} 
        {\rm SF_n}(\ell) = C_n \epsilon^{n/3} 
        \ell^{n/3 + \mu(n)}
        \label{eq:magKRSH}
    \end{equation}
    where $\mu(n)$ is called the intermittency parameter, and $\mu =0$ for Gaussian turbulence lacking intermittency. 

    In order to observe small scale intermittency, in the analysis below, it is important to compute diagnostics such as SDK at scales spanning the inertial range; in particular, scales closer to the dissipation range.
    SDK (see below) is related to magnetic increments as defined above.  
    One use of the second-order structure function ${\rm SF_2}$ is to help determine the inertial range for a given interval.  
    Ignoring intermittency, ($\mu(2) = 0$), K41 theory \citep{Kolmogorov1941} implies that for magnetohydrodynamic (MHD) fluids at high enough ${\rm R_e}$, we expect ${\rm SF_2}$ to scale as $\ell^{2/3}$ within the inertial range, outside of which ${\rm SF_2}$ will behave differently.
    
    Computing this quantity will help determine the lags within the inertial range for that particular interval.
    By doing so, we lay out the scales at which to compute SDK in order to properly examine intermittency.
    In Appendix \ref{app:SF}, Figure \ref{fig:SF2} suggests that spatial lags between $10~{\rm d_i}$ and $1000~{\rm d_i}$ lie reasonably within the inertial range for all three spacecraft.
    As will be discussed in Appendix \ref{app:data}, the time resolution for the individual spacecraft's magnetometer places a lower limit on the scale for measuring intermittency.
    Voyager 1 and PSP data were able to resolve down to $10~{\rm d_i}$, whereas Helios 1 data resolved down to $120~{\rm d_i}$.
    
%%%%%%%%%%%%%%%%%%%%%%%%%%%%%%%%%%%%%%%%%%
\subsection{Scale Dependent Kurtosis}\label{subsec:SDK}

    The scale dependent kurtosis is the ratio of the fourth statistical moment to the square of the second moment. For a scalar quantity with Gaussian distribution the ${\rm SDK} = 3$.
    Therefore for  increments of a scalar with Gaussian distributions for all lags, the SDK does not vary with lag, assuming the ensemble is stationary and otherwise well-behaved.
    If SDK takes on  values $> 3$, its distribution will have wide tails that deviate from a Gaussian distribution.
    Typically in turbulence, the distribution of velocity or magnetic field component increments \citep{SorrisoValvoEA1999} are Gaussian at large lags, and exhibit fat tails at small lags in the inertial range.  
    In such cases one expects the SDK to increase with decreasing scale in the inertial range. 
    Intermittency is associated with a super-Gaussian kurtosis, or $SDK>3$.
    
    The SDK of the $X$-component increments of  magnetic field fluctuations, $\Delta b_X = \hat X \cdot \Delta \Vec{b}$,
    is defined as, with $\ell=-V_{SW} \tau$:
    
    \begin{equation}\label{eq:SDK}
        \kappa_X(\ell) = \frac{\left< | \Delta b_X(t,\tau) |^4 \right>}{\left< | \Delta b_X(t,\tau) |^2 \right>^2}.
    \end{equation}
    A general expectation based on the refined similarity is that 
    
    \begin{equation}
        \kappa_X(\ell) =  \frac{C_4}{C_2^2} \ell^{\left( \mu(4)-2\mu(2) \right)}
        \label{kappakrsh}
    \end{equation}
    where we assume as in Eq. \eqref{eq:magKRSH} that
    refined similarity is applicable to magnetic increments.
    In this framework any dependence on Reynolds number remains unspecified. However Reynolds number variation of $\kappa$ may be inferred 
    from additional empirical and theoretical considerations, which we pursue below. 
    As a prediction based from our previous 
    observational results 
    \citep{ParasharEA2019}
    with Voyager 1, we expect stronger intermittency at smaller heliocentric distances (larger ${\rm R_e}$).
    
    We may also examine how the kurtosis at a fixed scale evolves with varying heliocentric distance.
    To formulate a theoretical description, we need to find a link between the kurtosis and the two scaling factors, ${\rm R_e}$ and lag $\ell$.
    To illustrate how intermittency and Reynolds number are intertwined in this way, we temporarily adopt a special case in which the intermittency statistics are represented by a log-normal distribution \citep{Obukhov1962}.
    \added{
    %{\bf
    In a later section we will introduce other intermittency theories \citep{She&Leveque1994,Marsch&Tu1997} for use in certain comparisons.
    %}
    }
    This specialization selects a particular Reynolds number dependence, which we may write explicitly, as follows. 
    Referring to \citet{VanAtta&Antonia1980}, their expression for the n-th order structure function is:
    
    \begin{equation}\label{eq:sflognorm}
        \left< \left( \Delta u \right)^n \right> =
        C_n
        \left[ \epsilon \ell \right]^{n/3}
        \left( \frac{\ell}{L} \right)^{\mu_0 n \left( 3-n \right)/18}
    \end{equation}
    where $L$ inherits the spatial definition of the correlation scale and $\mu_0$ is assumed to be a universal constant.
    Note that the log-normal scaling of the $n$-th order structure function in Eq. \eqref{eq:sflognorm} corresponds to the more general formulation in KRSH (see e.g,, \citet{ChhiberEA2021April}), as in Eq. \eqref{eq:magKRSH},
    where for this special case
    one identifies
    
    \begin{equation}\label{eq:muln}
        \mu(n) = \mu_0 n (3-n)/18.
    \end{equation}
    Further manipulating Eq. \eqref{eq:sflognorm}, we can eliminate $L$ in favor of ${\rm  R_e}$ and $\eta$, and approximate $\eta = {\rm d_i}$ as has been our practice; cf. Section \ref{subsec:Re}.
    Since $L \sim {\rm R_e}^{3/4} \cdot {\rm d_i}$, Equation \eqref{eq:sflognorm} may also be expressed in the form 
    
    \begin{equation}\label{eq:sfsub}
        \left< \left( \Delta u \right)^n \right> =
        C_n
        \left [ \epsilon \ell \right ]^{n/3}
        {\rm  R_e}^{-3 \mu(n) /4} \left( \frac{\ell}{\rm d_i} \right)^{\mu(n)}.
    \end{equation}
    As a result, we can write the kurtosis of the velocity increments as
    
    \begin{align}
        \frac{\left< \left( \Delta u \right)^4 \right>}{\left< \left( \Delta u \right)^2 \right>^2}
        & = \frac{C_4}{C_2^2} \frac{\left( \epsilon \ell \right)^{4/3} 
            {\rm R_e}^{-3\mu(4)/4} \left( \frac{\ell}{\rm d_i} \right)^{\mu(4)}}{
            \left[ \left( \epsilon \ell \right)^{2/3} 
            {\rm R_e}^{-3\mu(2)/4} \left( \frac{\ell}{\rm d_i} \right)^{\mu(2)} \right]^2} \label{eq:sdklogn} \\
        & \sim \left[ {\rm R_e}^{-3/4} \left( \frac{\ell}{\rm d_i} \right) \right]^{\mu(4)-2\mu(2)}\label{eq:sdkvelocity}
    \end{align}
    Note that \citet{VanAtta&Antonia1980} arrive at the conclusion that $\mu_0 \approx 0.25$ based on hydrodynamic experimental observations, a result that may not apply to magnetic field observations in space. 
    In fact, below we will find that the observed magnetic field intermittency statistics in the solar wind depart considerably from the Van Atta - Antonia result when computed as in this exercise employing log-normal statistics.
 
%%%%%%%%%%%%%%%%%%%%%%%%%%%%%%%%%%%%%%%%
%%%%%%%%%%%%%%%%%%%%%%%%%%%%%%%%%%%%%%%%
\section{Results}\label{sec:results}
%%%%%%%%%%%%%%%%%%%%%%%%%%%%%%%%%%%%%%%%

    The physical quantities needed to study effective Reynolds number and scale dependent kurtosis will be examined below from datasets obtained from three missions - Voyager 1, Helios 1, and PSP. 
    Taken together the subsets of the associated datasets that we employ span heliocentric distances from $0.16~{\rm au}$ to $10~{\rm au}$.
    Details of the data are provided in Appendix \ref{app:data}.
    \added{
    %{\bf 
    For all power-law fittings in Section \ref{sec:results} and in the Appendix, their corresponding uncertainty is provided explicitly. 
    For the figures that have error bars, the limits on those bars are fixed to its corresponding uncertainty.
    All uncertainties are attained by using the standard error (goodness-of-fit), a percentage of the fitted power-law scaling representing the average deviation of any point from the fitted power-law.
    %}
    }

\subsection{Correlation and Ion Inertial Length Radial Scaling} \label{sec:corr_di}

    First we compute the auto-correlation of the Cartesian (RTN) magnetic field fluctuations to determine the correlation scale.  
    Referring to Section \ref{subsec:corr}, the correlation time $\tau_e$ corresponds to the condition
    $R_C(\tau_e)=1/e$.
    We then apply the Taylor hypothesis to convert $\tau_e$ to $\lambda_C$.  
    \begin{figure}[htp!]
        \centering
        \includegraphics[scale=.5]{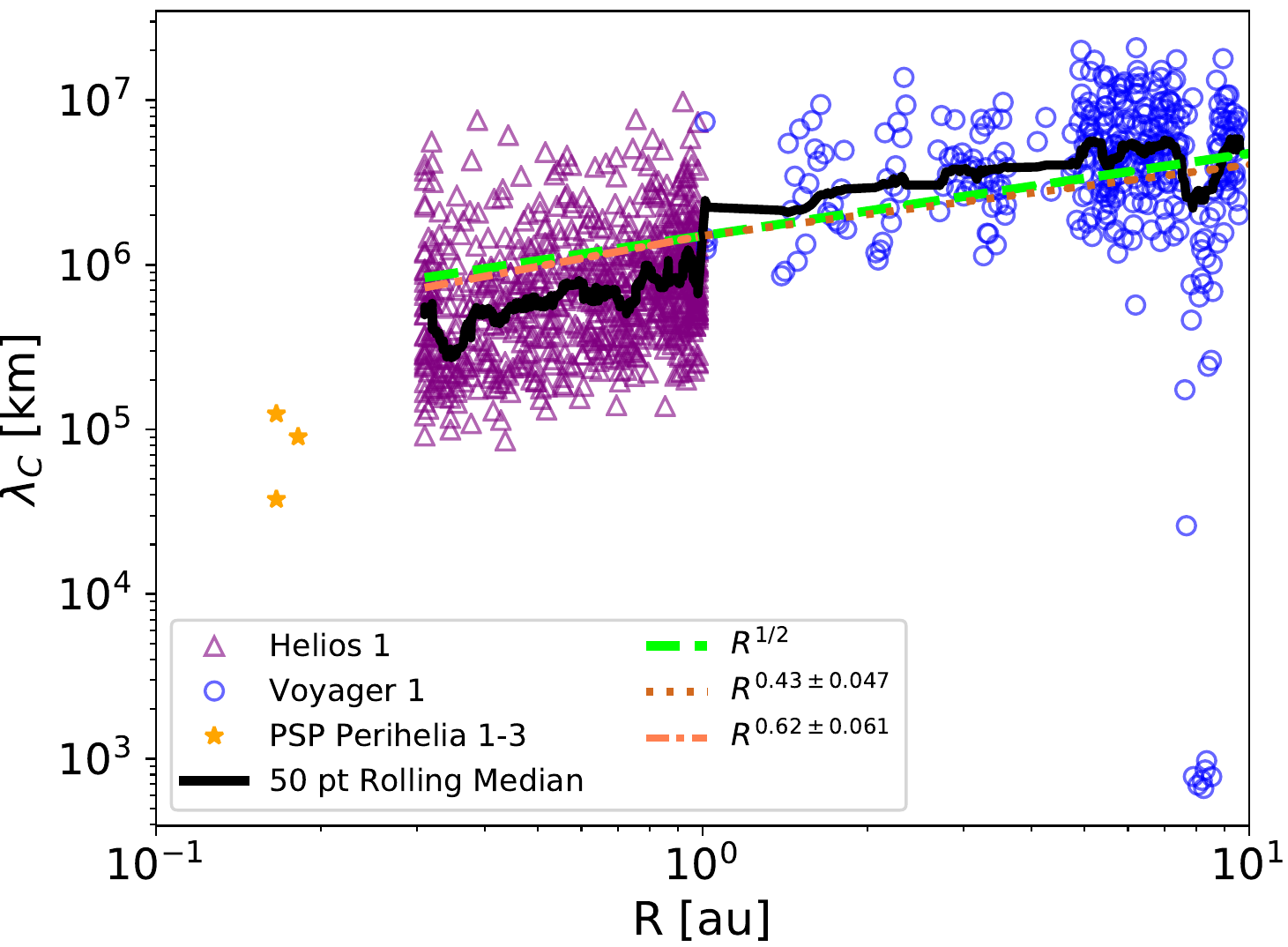}
        \caption{Combined correlation lengths ($\lambda_C$) for PSP, Helios 1, and Voyager 1 magnetic-field measurements, as a function of heliocentric distance.  A 50-point rolling median is plotted along with a reference $R^{1/2}$ power-law (green dashed line).  Power-law fits of $R^{0.62}$ for Helios and $R^{0.43}$ for Voyager are given with uncertainty $0.061$ and $0.047$, respectively.}
        \label{fig:CL}
    \end{figure}
    As shown in Figure \ref{fig:CL}, we observe that the radial trend of the magnetic field correlation scale $\lambda_C$, for both Voyager and Helios data, is nearly consistent with a power-law of $\sqrt{R}$ as discussed in Section \ref{subsec:corr}.
    \added{
    %{\bf
    Helios shows a slightly stronger radial dependence of $R^{0.62 \pm 0.061}$ whereas Voyager shows a slightly weaker power-law of $R^{0.43 \pm 0.047}$.
    %}
    }
    There is, however, considerable scatter.
    We must keep in mind that most solar wind parameters are quite variable, and the spatial correlation scale may be expected to vary due to a number of physical and methodological factors.
    
    There is however an apparent decrease of correlation lengths at $\approx 0.16~{\rm au}$.
    This is associated with the PSP samples.
    A more detailed analysis would be required to determine if this is a statistically significant decrease.
    We should note that a corresponding decrease of the correlation length is not found in global MHD simulations \citep{UsmanovEA2018}, which compute correlation lengths as part of an incorporated subgridscale turbulence model that generally agrees approximately well with PSP observations \citep{ChhiberEA2021Dec}.
    
    Another possibility that may explain the decrease in the radial trend is that near perihelion, PSP is usually measuring correlation lengths parallel to the large scale magnetic field, and these may be systematically smaller than the more perpendicular correlation scales measured at larger heliocentric distances.
    This possibility is supported by a previous study using Helios data \citep{RuizEA2011}.
    The finding was that correlation lengths measured parallel to the mean magnetic field are systematically smaller than perpendicular correlation scales in the inner orbits of Helios.
    This disparity was found to be diminished moving outward towards larger heliocentric distances.
    It is discussed further below in Section \ref{sec:conclusions}, but for now we note the small correlation scale values at PSP, compared to the trend line from other spacecraft recorded at greater distances, may be in part a systematic selection effect.
    
    \replaced{\sout{Also, there are noticeably two additional regions where one finds significantly smaller values of $\lambda_C$ over relatively small distance ranges --  at distances $R \approx 4.5~{\rm au}$ and $R \approx 8.5~{\rm au}$.
    For $R \approx 4.5~{\rm au}$, Voyager is encountering Jupiter's magnetosphere, which is inherently large in size ($> 1~{\rm au}$ from bow shock to magnetotail).
    Although the data corresponding to Voyager's nearest approach to Jupiter was removed, the shortening of $\lambda_C$ may be a remaining direct result of Jupiter's magnetosphere.
    In the vicinity of $R\approx 8.5~{\rm au}$, the observed cluster of shortened values of $\lambda_C$ may be a result of encountering pickup ions.
    For each of these cases, we decided to ignore these regions in results involving ${\rm R_e}$ since they strongly deviate from the average trend.}}
    {
    %{\bf 
    There is another region where one finds significantly smaller values of $\lambda_C$ over relatively small distance ranges --  at distance $R \approx 8.5~{\rm au}$.
    There may be two possible reasons for this observed effect.
    The observed cluster of shortened values of $\lambda_C$ may be a result of encountering pickup ions (PUIs), which can inject energy at high wave numbers in the inertial range, pulling the correlation scale toward smaller scales \citep{ZankEA1996,BreechEA2008}.
    It is also possible that a combination of Jovian and Saturnian magnetospheric effects contributed to this decrease in $\lambda_C$.
    Jupiter's magnetosphere is inherently large extending out to regions near that of Saturn's orbit \citep{KhuranaEA2004}, and there is evidence that the Voyagers' trajectory passed through Jupiter's magnetotail near $8~{\rm au}$ \citep{ScarfEA1981}.
    The same effect was seen in \citet{ParasharEA2019}, however, here we do not observe shortened values of $\lambda_C$ at distance $R \approx 4.5~{\rm au}$ as previously observed.
    This is the result of a change in both interval size and statistical criterion, compared to that used in \citet{ParasharEA2019} (for interval size and computational criterion, see Appendix \ref{subapp:comp}).
    Therefore, we ignore results in the region around $8.5~{\rm au}$ while fitting radial power-laws to ${\rm R_e}$ since they strongly deviate from the average trend.
    %}
    }
    
    The second length scale of interest is the ion inertial length ${\rm d_i}$.
    Its definition, Equation \eqref{eq:di}, only depends on the average proton density in a given interval.
    \begin{figure}[htp!]
        \centering
        \includegraphics[scale=.5]{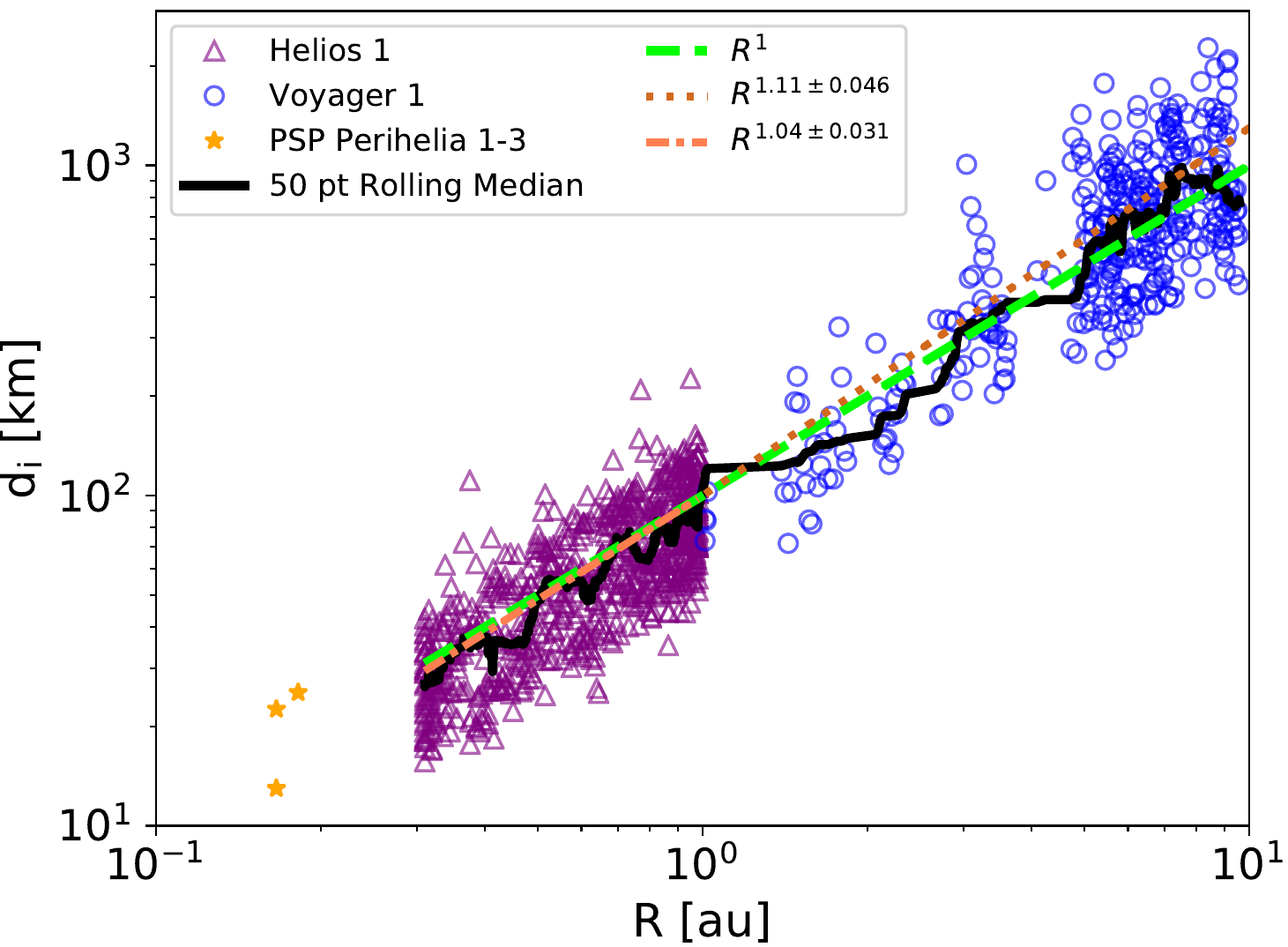}
        \caption{PSP, Helios 1, and Voyager 1 ion inertial lengths [km] at varying heliocentric distances, $R$ [au].  Overall trend is presented by a $50$ point rolling median, which follows very closely a $R^1$ power-law shown as a reference line (green dashed line). Power-law fits of $R^{1.04}$ for Helios and $R^{1.11}$ for Voyager are given with uncertainty $0.031$ and $0.046$, respectively.}
        \label{fig:Di}
    \end{figure}
    As depicted in Figure \ref{fig:Di}, the trend of the ion inertial lengths nearly scale linearly with $R$, a consequence of the $R^{-2}$ proton density radial scaling.
    The power-law fits for Helios and Voyager are $R^{1.04 \pm 0.031}$ and $R^{1.11 \pm 0.046}$, respectively.
    The general linear radial scaling works well even down to $\approx 0.16~{\rm au}$.

%%%%%%%%%%%%%%%%%%%%%%%%%%%%%%%%%%%%%%%%
\subsection{Effective Reynolds Number ${\rm R_e}$ \label{sec:Reff}}

\begin{figure}[htp!]
        \centering
        \includegraphics[scale=.5]{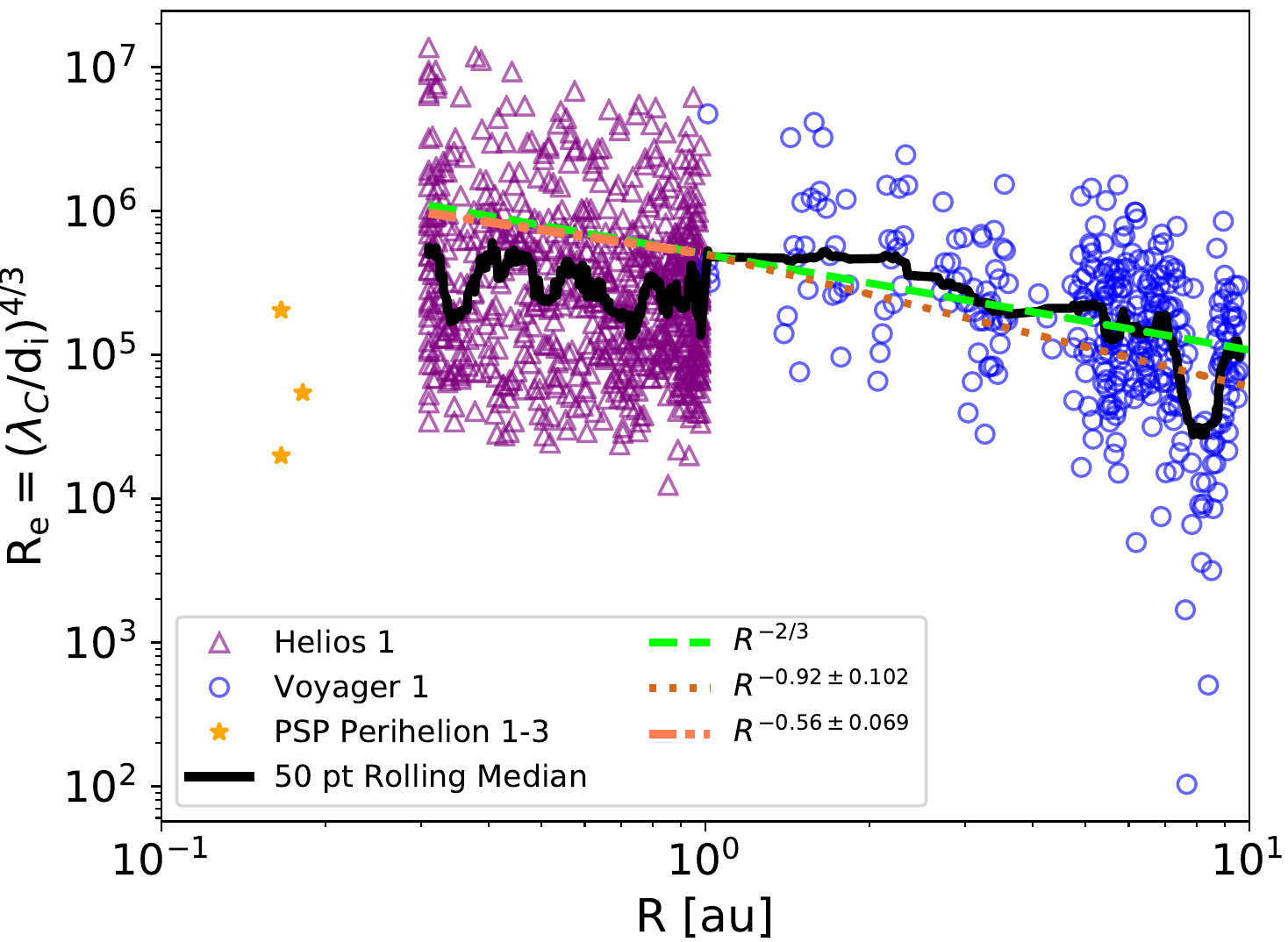}
        \caption{PSP, Helios 1, and Voyager 1 estimated effective Reynolds Number (${\rm R_e}$) as a function of heliocentric distance in au.  A $50$ point rolling median, the black solid line, is plotted over Helios 1 and Voyager 1 distributions , in addition to an unfitted $R^{-2/3}$ power-law (green dashed line). Power-law fits of $R^{-0.56}$ for Helios and $R^{-0.92}$ for Voyager are given with uncertainty $0.069$ and $0.102$, respectively.}
        \label{fig:Re}
    \end{figure}

    Using Eq. \eqref{eq:Re}, as well as the correlation lengths and ion inertial lengths measured in the Voyager, Helios and PSP datasets, we accumulate the estimates of effective Reynolds number; these are portrayed in Fig. \ref{fig:Re}. 
    \replaced{\sout{It is apparent that the radial scaling of ${\rm R_e}$ for both the Helios 1 and Voyager 1 analyses are well-behaved, following similar trends and with similar statistical spread. 
    This follows from the analogous regular behavior of both correlation and ion inertial lengths for these two datasets.
    Consequently, it is possible to affirm the predicted $R^{-2/3}$ scaling in a coarse-grained sense, as shown in Figure \ref{fig:Re}.}}
    {
    %{\bf
    The Helios radial scaling of $R^{-0.56 \pm 0.069}$ follows more closely to the predicted $R^{-2/3}$ scaling than the Voyager scaling of $R^{-0.92 \pm 0.102}$.
    Although there is significant spread propagated by the spreads of ${\rm d_i}(R)$ and $\lambda_C(R)$, the average trend points to decreasing ${\rm R_e}$ with increasing distance $R>0.30~{\rm au}$.
    %}
    }

    In contrast, the PSP Reynolds number data seems to be smaller than the average trend and strongly deviates from the $R^{-2/3}$ power-law, even taking into account the small sample.
    This is not associated with the behavior of density, as there is no apparent deviation from the ion inertial length radial scaling for PSP as seen in Fig. \ref{fig:Di}.
    As a result, the observed deviation of the radial trend of correlation lengths for the PSP perihelia are responsible for the observed drop in ${\rm R_e}$ at those distances in Fig. \ref{fig:Re}. 
    Possible causes for this are discussed in the previous section, and further discussed in Section \ref{sec:conclusions}. 
    
\subsection{Scale Dependent Kurtosis (SDK)} \label{sec:SDK}
 
\begin{figure}[htp!]
        \centering
        \includegraphics[scale=.6]{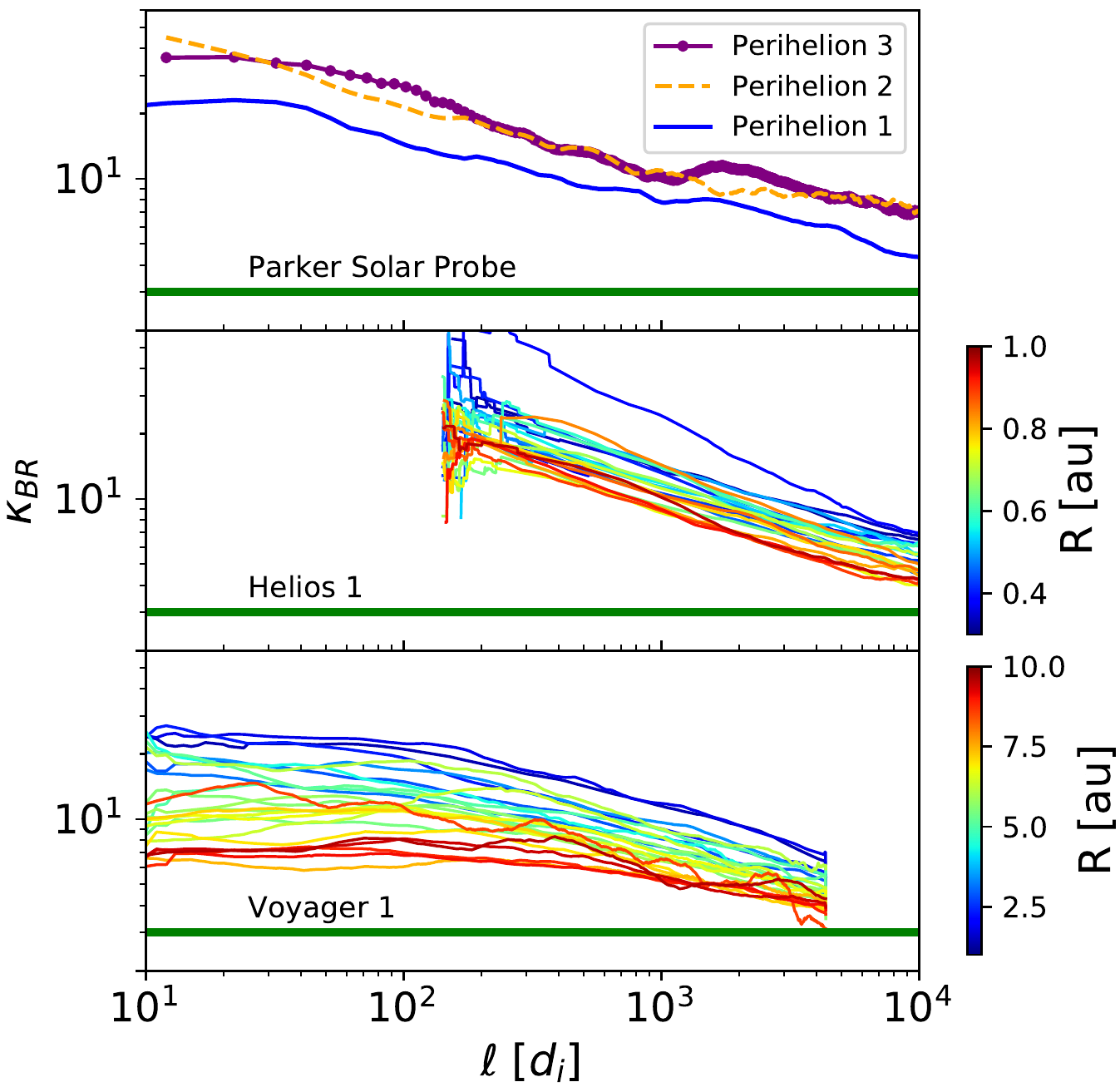}
        \caption{SDK for PSP (top), Helios 1 (middle), and Voyager 1 (bottom).
                    Curves for Helios and Voyager represent averages of SDK over 15 nearby available intervals with their colors keyed to the average heliocentric distance. The solid horizontal green line marks $SDK=3$.}
        \label{fig:sdkbr}
    \end{figure}

    In this section, we examine how the kurtosis measured in the solar wind datasets varies with lag and with heliocentric distance.  
    As noted in Section \ref{subsec:SDK}, the typical non-Gaussian probability distributions with ``fat tails'' found in solar wind increments at smaller inertial range lags \citep{SorrisoValvoEA1999} have SDK greater than $3$.
    The kurtosis of longitudinal magnetic field increments computed from the PSP, Helios and Voyager datasets are shown in the respective panels of Figure \ref{fig:sdkbr}.
    Each panel of Figure \ref{fig:sdkbr} shows estimates of scale dependent kurtosis computed as a function of lag in units of ion inertial scale.
    For clarity, each curve that is shown, for Helios and Voyager, is an average of 15 cases computed from nearby available intervals. 
    For the Voyager and Helios panels, the color of the curve is keyed to the heliocentric distance. 
    
    Figure \ref{fig:sdkbr} demonstrates a consistent picture that SDK is larger at smaller scale, as is expected in general. It also shows, in the Helios (middle panel of Fig. \ref{fig:sdkbr}) and Voyager (bottom panel of Fig. \ref{fig:sdkbr}) data, that the kurtosis at a given normalized scale decreases with increasing heliocentric distance, with only a few exceptions. The general trend is towards weaker intermittency at larger $R$.
    
    \replaced{\sout{ There are two regions where the kurtosis deviates from the overall trend observed by Voyager 1.
    The first instance occurs near Voyager 1 encounters Jupiter, and may be due to Jupiter's upstream waves and other foreshock activity \citep{SmithEA1983}. A clear indication is seen in the decrease of kurtosis values near $5~{\rm au}$ in Fig. \ref{fig:SDKfixdi}. 
    A likely effect is that those waves result in magnetic field fluctuations that are 
    uncorrelated in phase, and thus responsible for a tendency towards Gaussian kurtosis (see discussion in \citet{HadaEA2003,WanEA2012}).
    The second event occurs around $8~{\rm or}~{ 9}~{\rm au}$, which we suspect to be a result of pickup ions.}}
    {
    %{\bf
    There is one region where the kurtosis deviates from the overall trend observed by Voyager 1 -- at distance $R \approx 8.5~{\rm au}$ -- that can be clearly indicated in the decrease of kurtosis values in Fig. \ref{fig:SDKfixdi}.
    This decrease can be produced by wave activity associated with Jupiter's magnetotail \citep{ScarfEA1981}.
    Such wave activity can result in magnetic field fluctuations that are uncorrelated in phase \citep{SmithEA1983}, and thus responsible for a tendency towards Gaussian kurtosis (see discussion in \citet{HadaEA2003,WanEA2012}).
    The local magnetic fluctuations from pickup ions can also produce the same effect.
    %}
    }
    
    This feature in the kurtosis occur at ranges of heliocentric distance for which the correlation scale is seen to shorten relative to the prevailing trend (c.f. Section \ref{sec:corr_di}).
    Our working hypothesis is that these features are explained in both cases by injection of incoherent fluctuations by wave-particle interactions.
    However, apart from this specific region, intermittency generally weakens with both increasing heliocentric distance and scale.
    Power-law analysis for the inertial range is discussed later in Section \ref{sec:mu}.
    Additional information for the SDK of the tangential and normal magnetic field increments can be found in Appendix \ref{app:SDK}.
    
    To provide a quantitative context for these observations, we may adopt two key assumptions -- the validity of the refined similarity hypothesis, and a log-normal distribution of increments.
    This allows us to write the SDK of velocity increments as a function of lag $\ell$ and ${\rm R_e}$ as in Eq. \eqref{eq:sdkvelocity}.
    Below we will exploit this approach to separately examine the variation of kurtosis with properly normalized physical scale, and with effective Reynolds number ${\rm R_e}$.
    These details are presented in the next three subsections.
    In selecting scales of interest, we focus on the inertial range, which can be identified (see Figure \ref{fig:SF2}) as the range in which SF$_2 \sim \ell^{2/3}$.
    For a fixed ${\rm R_e}$, then the kurtosis of the longitudinal velocity increments is expected to scale as $\kappa \sim \ell^{\mu(4)-2\mu(2)}$ \citep{ChhiberEA2021April}.
    
\subsection{Radial Variation of SDK at Constant Multiple of ${\rm d_i}$}
\label{sec:sdkdi}

\begin{figure}[htp!]
        \centering
        \includegraphics[scale=.5]{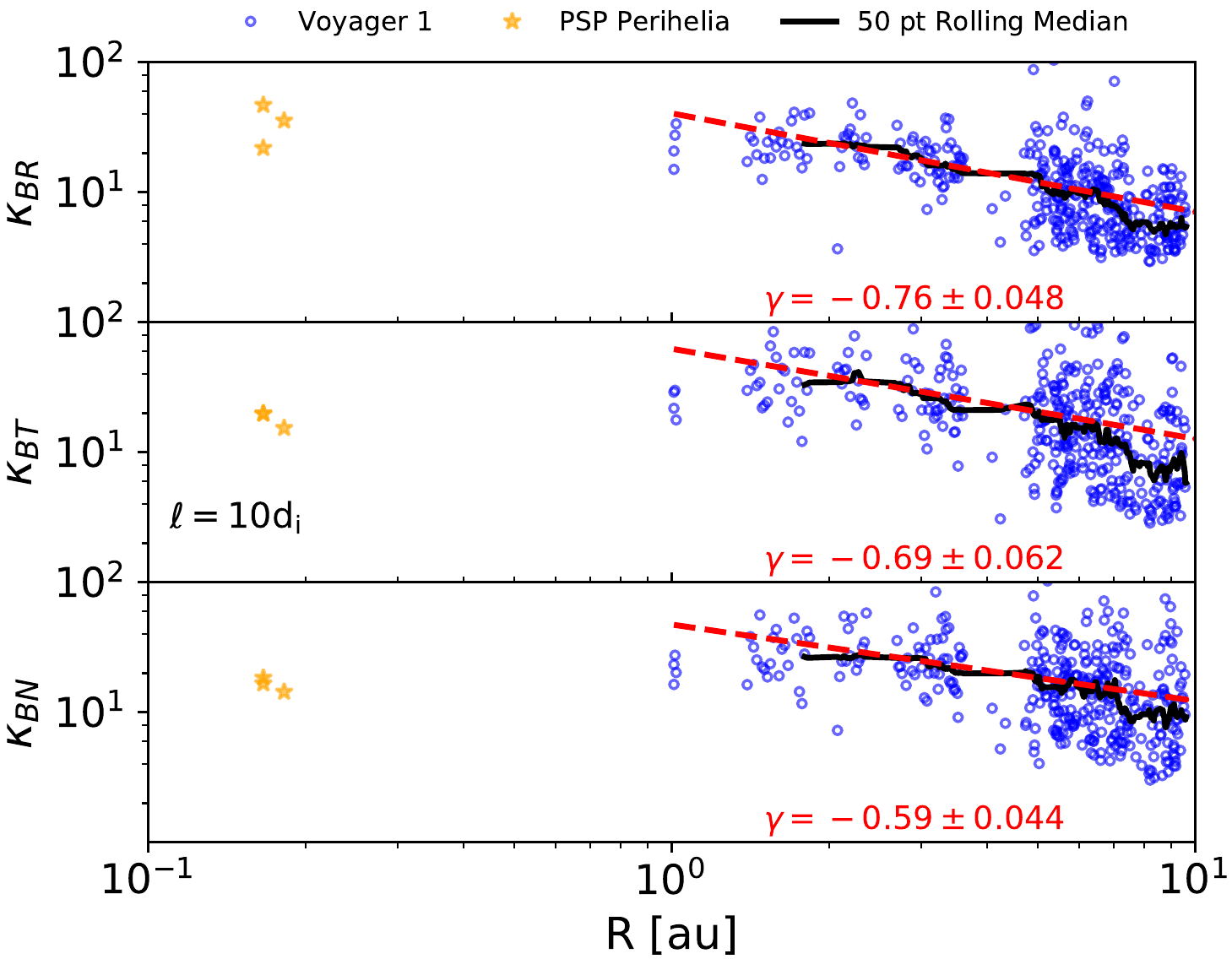}
        \vskip0.2in
        \includegraphics[scale=.5]{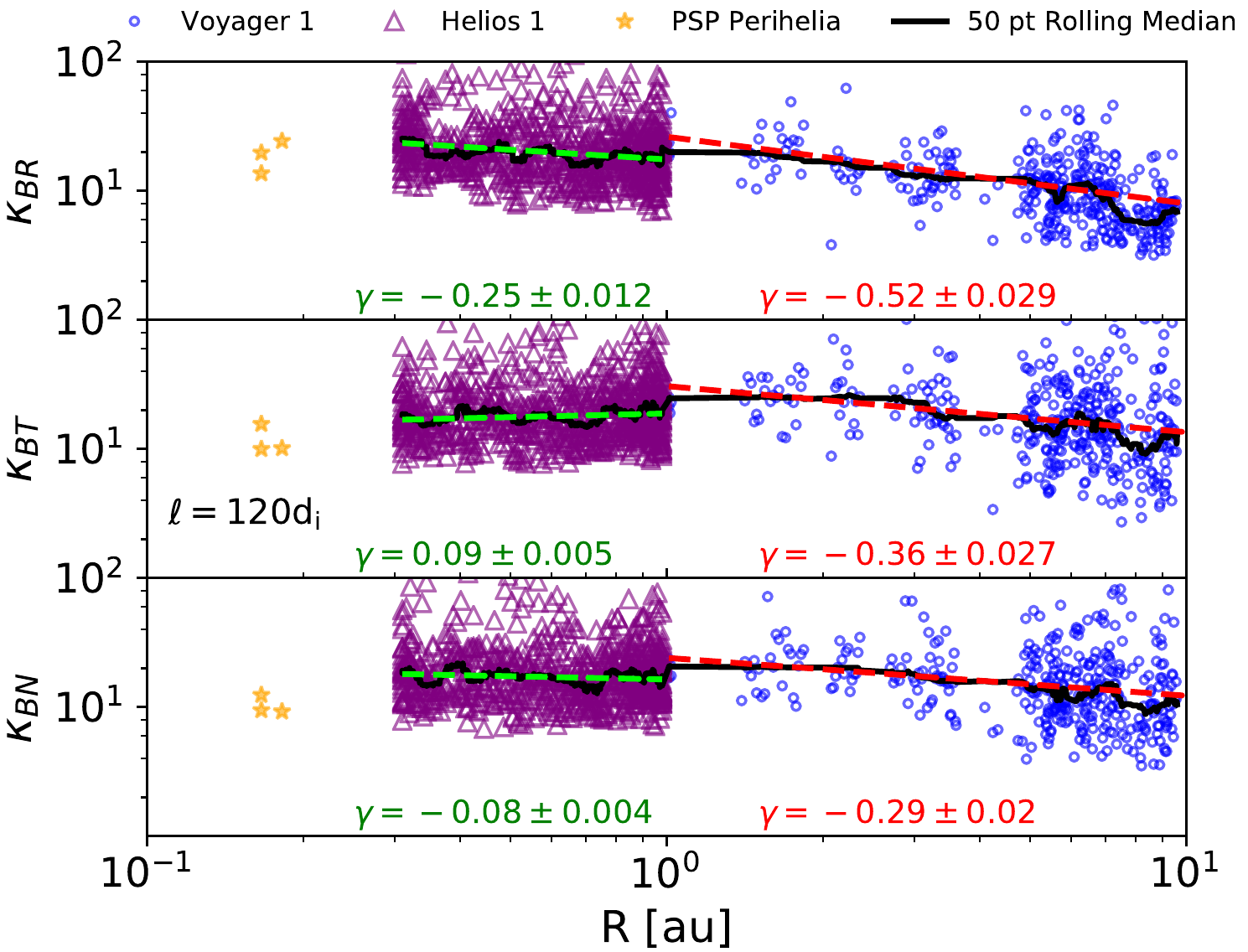}
        \caption{PSP, Helios 1, and Voyager 1 radial variation of RTN components of $\kappa$ fixed at $10~{\rm d_i}$ (top) and $120~{\rm d_i}$ (bottom).  Information for power-law fits for Helios 1 and Voyager 1, separately, are shown under the dashed curve in the same color, respectively. $\gamma$ represents the fitted power-law dependence with radius
        of the SDK at a fixed lag $\ell$. Uncertainties are provided explicitly for the corresponding fitted values of $\gamma$.}
        \label{fig:SDKfixdi}
    \end{figure}

    Instead of holding ${\rm R_e}$ constant and varying $\ell$, now we hold $\ell$ constant after proper normalization, and examine SDK behavior as a function of heliocentric distance.
    First, we look at lags that are multiples of the inner scale, specifically $\kappa_{BR}(\ell=10~{\rm d_i})$ for PSP and Voyager 1, and $\kappa_{BR}(\ell=120~{\rm d_i})$ for all three spacecraft.
    The latter choice is a result of the limited Helios 1 data resolution that cannot resolve down to $10~{\rm d_i}$.
    We now describe results for $\kappa$ at these fixed physical scales 
    vs heliocentric distance
    $R$ 
    and provide power-law fits to the radial trend as 
    $\kappa \sim R^\gamma$.
    \added{
    %{\bf
    Note that here we are assuming that the kurtosis depends on $R$ as a strict power-law; however, this is only in an attempt to extract a general trend of the kurtosis as a function of distance.
    In fact, it is clear from Fig. \ref{fig:SDKfixdi} that a single power-law will not accurately describe the radial evolution of the kurtosis over the entire range of heliocentric distance.
    %}
    }

\begin{figure}[htp!]
        \centering
        \includegraphics[scale=.5]{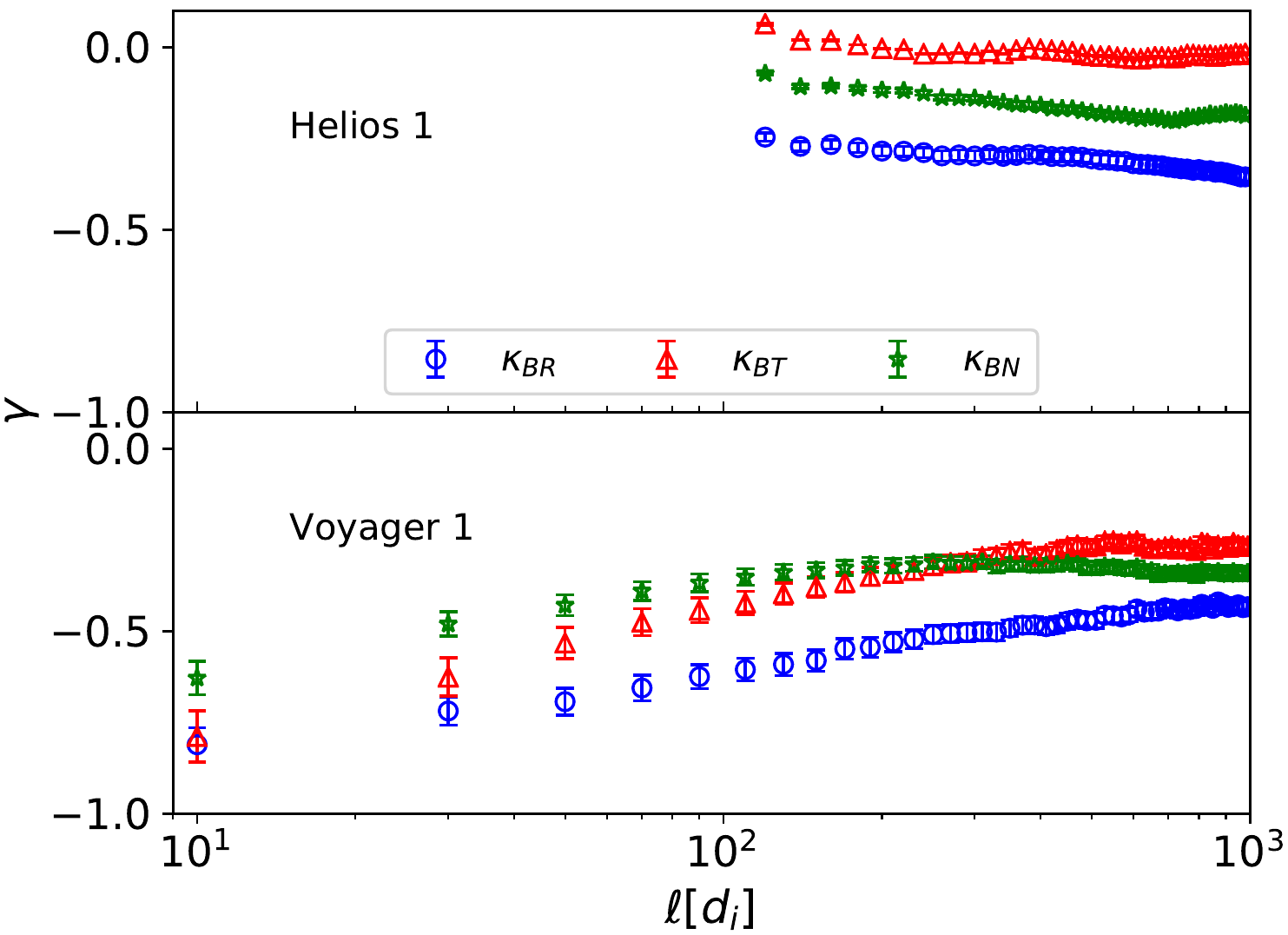}
        \caption{Helios 1 (top) and Voyager 1 (bottom) $\gamma$ as a function of lag $\ell$ calculated for each component of the magnetic kurtosis. Note that $\gamma$ refers to the fitted parameter for the radial dependence of $\kappa(\ell/{\rm d_i}) \sim R^\gamma$, with error bars associated with the corresponding uncertainty in $\gamma$.}
        \label{fig:gamma}
    \end{figure}

    The top panel of Figure \ref{fig:SDKfixdi} shows the Voyager and PSP results for $\kappa$ at $10~{\rm d_i}$, for Cartesian magnetic field components R,T, and N.
    Weaker intermittency is found at larger heliocentric distance $R$ at this scale, excluding PSP (see discussion in Section \ref{sec:conclusions}).
    This is supported by the value of $\gamma$ at varying lags, which is nearly negative for all Cartesian components of the fluctuating magnetic field (see Fig. \ref{fig:gamma}).
    The uncertainty of $\gamma$ is given explicitly in Figure \ref{fig:SDKfixdi}.
    
    However, for the larger lag $\ell=120~{\rm d_i}$, we report in the bottom panel of Fig. \ref{fig:SDKfixdi}
    slightly different radial trends when comparing Helios and Voyager observations.
    On average, the tangential and normal kurtosis components are weakly dependent on $R$ across PSP and Helios results, with a transition to a stronger dependence for Voyager results.
    The tangential kurtosis for Helios shows a slight increasing trend for increasing heliocentric distance, at which point $\gamma$ for Voyager is negative for distance $R>1~{\rm au}$.
    On the contrary, the kurtosis of the radial magnetic field increment, which is based on the standard longitudinal magnetic field component, shows a clear trend toward less intermittency at larger $R$.
    \deleted{
    %{\bf
    Additionally, as one increases the scale, all components of kurtosis from Voyager data with increasing heliocentric distance; whereas the radial kurtosis continues to decrease but at a slower rate compared to the kurtosis held at the smaller lag $\ell=10~{\rm d_i}$.
    %}
    }
    
    We also investigate further into how $\gamma$ is dependent on $\ell$.
    In Figure \ref{fig:gamma}, $\gamma$ computed from all three components of Helios $\kappa$ are generally invariant with respect to $\ell$.
    However, Helios $\gamma$ values do slightly decrease with increasing $\ell$ and the tangential kurtosis even switches sign.
    This suggests that larger scale intermittency falls off faster than smaller scale intermittency as a function of $R$ for $0.30~{\rm au} < R < 1~{\rm au}$.
    On the contrary, $\gamma$ from all components of Voyager $\kappa$ suggest that small scale intermittency weakens faster than large scale intermittency for $1~{\rm au} < R < 10~{\rm au}$.

\subsection{Radial Variation of SDK at Constant Fraction of $\lambda_C$}
\label{sec:sdkCL}
    
    We also investigate the kurtosis at scales that are constant fractions of $\lambda_C$.
    As expected, the dependence of SDK on heliocentric distance changes accordingly.
    The corresponding expectation may be examined by letting $\ell = \chi \lambda_C$ where $\chi$, a number generally $<1$, is to be held constant.
    Inserting this in the right-hand side of Equation \eqref{eq:sdkvelocity}, which assumes a log-normal distribution,
    one finds that $\kappa \sim R^{\gamma} \left( \lambda_C/{\rm d_i} \right)^{2\gamma} \sim R^0$, the latter step following from the approximation discussed earlier that $\lambda_C(R)/{\rm d_i}(R) \sim R^{-1/2}$.
    In this case the kurtosis of the radial velocity field increments would not vary with radial distance $R$.
    
\begin{figure}[htp!]
        \centering
        \includegraphics[scale=.5]{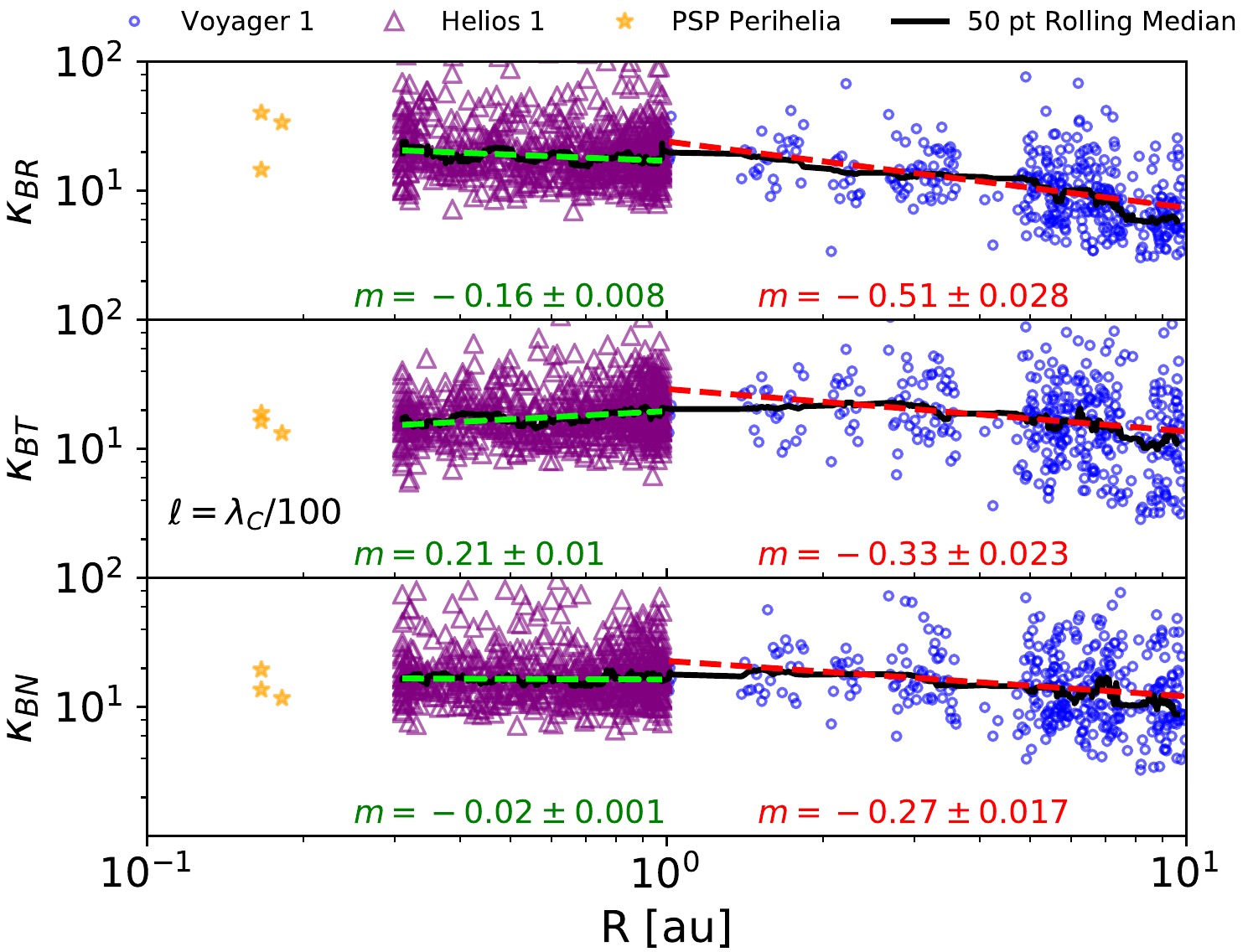}
        \vskip0.2in
        \includegraphics[scale=.5]{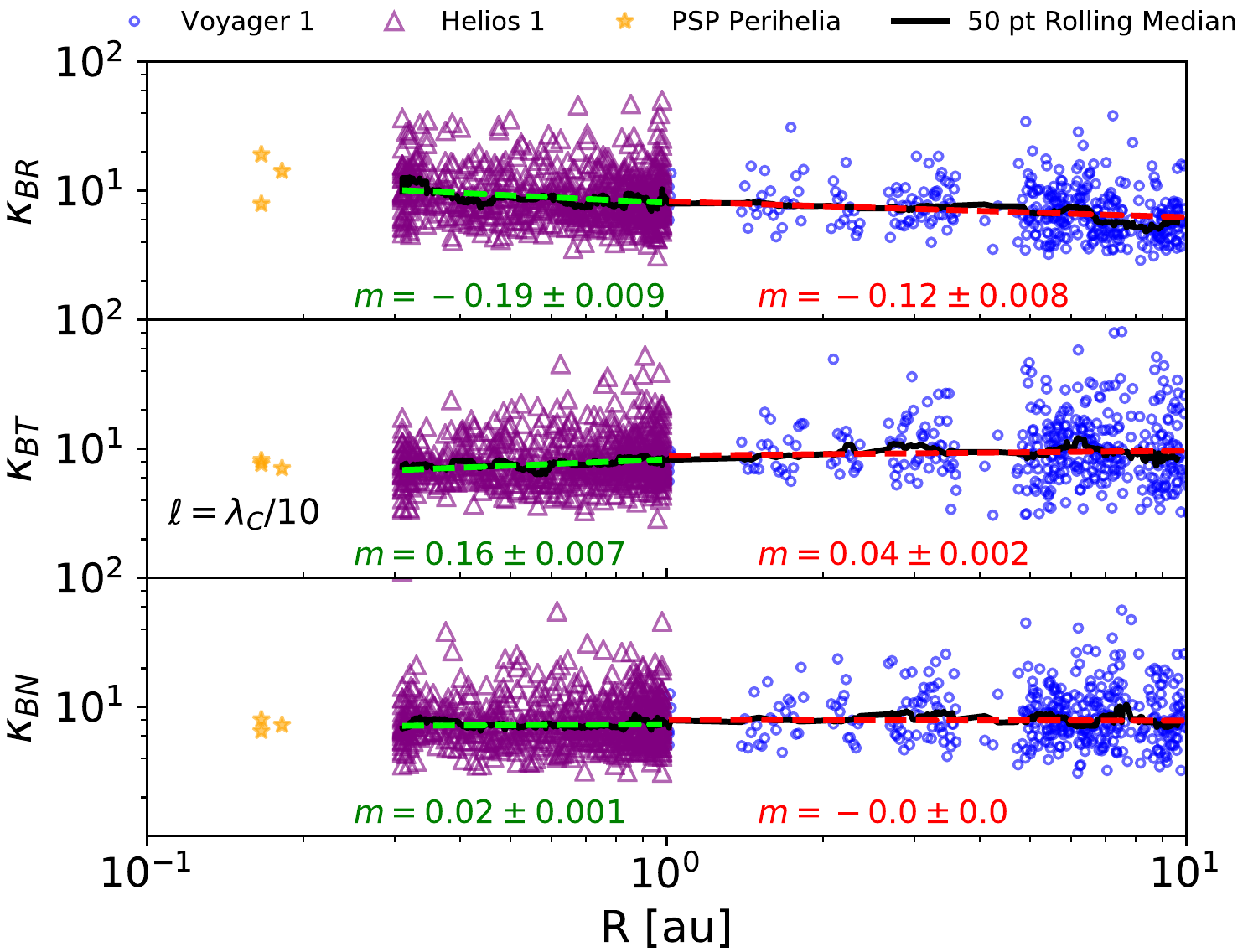}
        \caption{PSP, Helios 1, and Voyager 1 radial variation of RTN components of $\kappa$ fixed at $\lambda_C/100$ (top) and $\lambda_c/10$ (bottom). Power-law fits for Helios 1 and Voyager 1, separately, are shown under the dashed curve as in Figure \ref{fig:SDKfixdi}.  $m$ represents the assumed power-law dependence of the SDK on heliocentric distance $R$ at a fixed lag $\ell$.  Uncertainties are provided explicitly for the corresponding fitted values of $m$.}
        \label{fig:SDKfixL}
    \end{figure}

    To examine this case in the observations, we carry out the analysis using $\ell =\lambda_C/10$ and $\ell = \lambda_C/100$. 
    These results are shown in Fig. \ref{fig:SDKfixL}.
    We compute power-law fits to characterize the results,
    indicating the behavior as $\kappa \sim R^m$ in Figure \ref{fig:SDKfixL}.

    For $\ell=\lambda_C/10$, the bottom panel of Figure \ref{fig:SDKfixL} shows that the strength of intermittency on average remains approximately weakly dependent at all $R$.
    This is consistent with the predicted lack of dependence on heliocentric distance.
    At smaller scale $\ell = \lambda_C/100$, the kurtosis trend for Helios remains nearly the same as in the case for $\ell = \lambda_C/10$, decreasing slightly in value as shown in the top panel of Figure \ref{fig:SDKfixL}.
    However, there is a significant change of $m$ for all three components of Voyager $\kappa$, such that there is a weaker radial dependence of $\kappa$ for $\ell = \lambda_C/10$ compared to $\ell = \lambda_C/100$.
    These results suggest that the intermittency at scales related, and closer, to the outer scale is only weakly dependent on $R$.
    
    The results of this section are consistent with a heuristic picture of the emergence of intermittency as the cascade towards smaller scales progresses \citep{Matthaeus2021}. 
    Beginning with couplings at the energy containing scales, one may view the dominant quadratic nonlinearities in the MHD equation as affecting the transfer of energy from wavenumber $k$ to wavenumbers $q \sim 2k$. 
    If the statistical distribution of the originating modes is Gaussian, then it can be shown that a quadratic product, such that they enter the time derivatives, will have an exponential-like distribution.  
    The cascade, being quasi-local in scale, may be viewed as a succession of such couplings. 
    Thus as the cascade progresses and excites higher wavenumbers, these will have increasing non-Gaussianity. 
    That is, higher wavenumbers will tend to have stronger intermittency due to the greater number of factor of two ``hops'' needed to reach that part of the spectrum. 
    For $\ell$ a fixed fraction of $\lambda_C$, the number of hops to reach $\ell$ from the energy containing range is the same regardless of the Reynolds number. 
    Thus the kurtosis at fixed $\ell/\lambda_C$ should be insensitive to Reynolds number, consistent with our findings.  
    Conversely,
    at $\ell$ a fixed multiple of the inner scale ${\rm d_i}$, there will be more hops required to reach that scale for systems with larger inertial range bandwidth. 
    This implies that when ${\rm R_e}$ is increased, one should find higher $\kappa$ at fixed $\ell/{\rm d_i}$. 
    We now examine this expectation in the observations.  
    
\subsection{Dependence of kurtosis on effective Reynolds number}
\label{sec:sdkRe}

    The results of Section \ref{sec:Reff} demonstrate that on average the effective Reynolds number
    of interplanetary turbulence systematically changes (and in fact decreases)
    with increasing heliocentric distance $R$. 
    Therefore we may examine the Reynolds number  dependence of the kurtosis by evaluating its variation with $R$. To facilitate 
    meaningful comparison with theory,
    the evaluation of 
    the kurtosis proceeds 
    at a fixed physical scale, here 
    a fixed $\ell/{\rm d_i}$. 
   
    It is important to emphasize that the refined similarity hypothesis itself leads to the expression for $\kappa(\ell)$ in Eq. \eqref{kappakrsh}, from which we cannot anticipate variations with Reynolds number, as ${\rm  R_e}$ does not appear explicitly.
    (We note that in refined similarity there is residual, presumably weak implicit variation with ${\rm  R_e}$ in the constants $C_4$ and $C_2$ which we ignore here.)
    However if we adopt the log-normal hypothesis \citep{VanAtta&Antonia1980} as an approximation, then the relevant equation is Eq. \eqref{eq:sdkvelocity}, and in that expression ${\rm  R_e}$ appears explicitly. 
    Accordingly, we may proceed as follows.  
    
\begin{figure}[htp!]
        \centering
        \includegraphics[scale=.5]{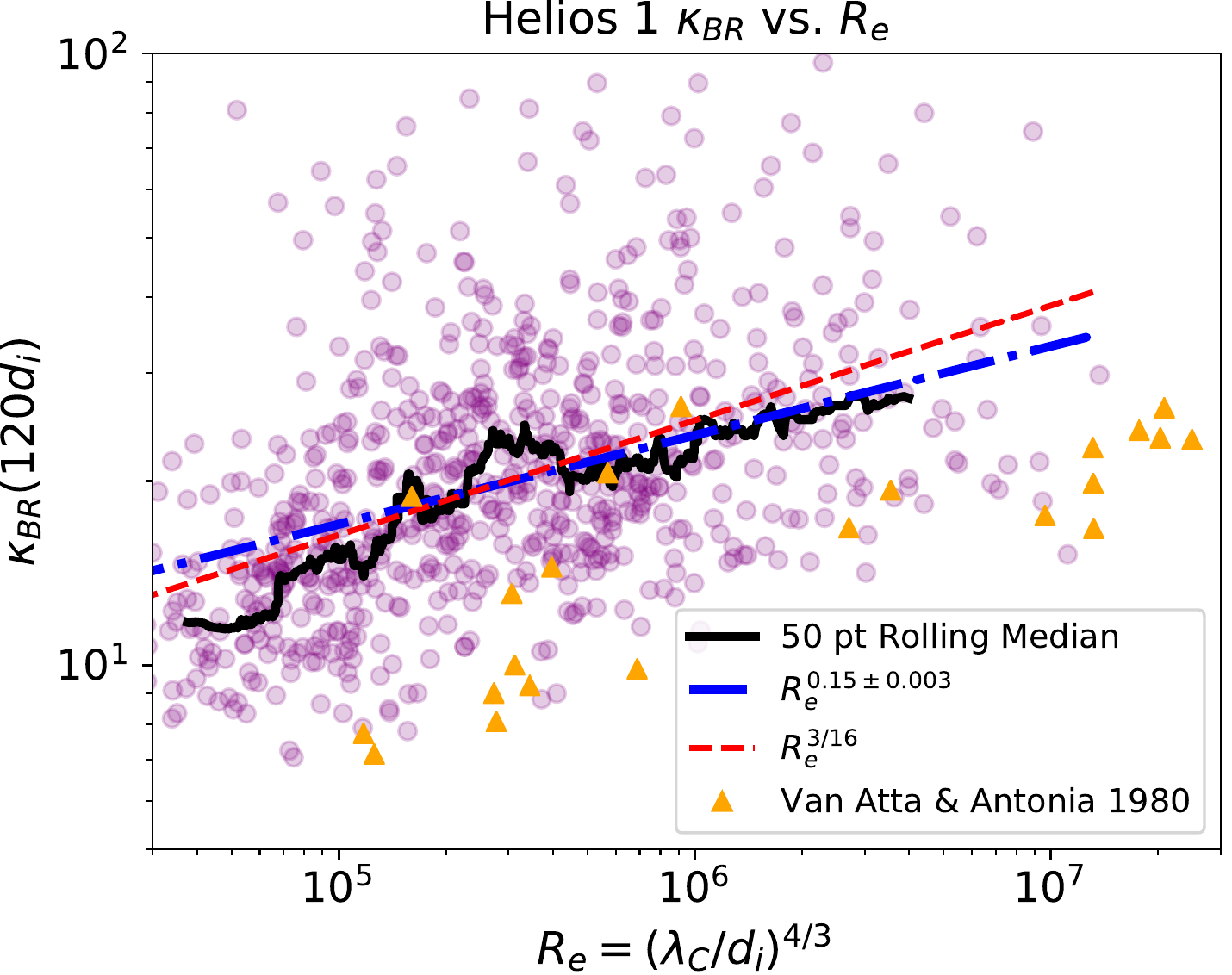}
        \vskip0.2in
        \includegraphics[scale=.5]{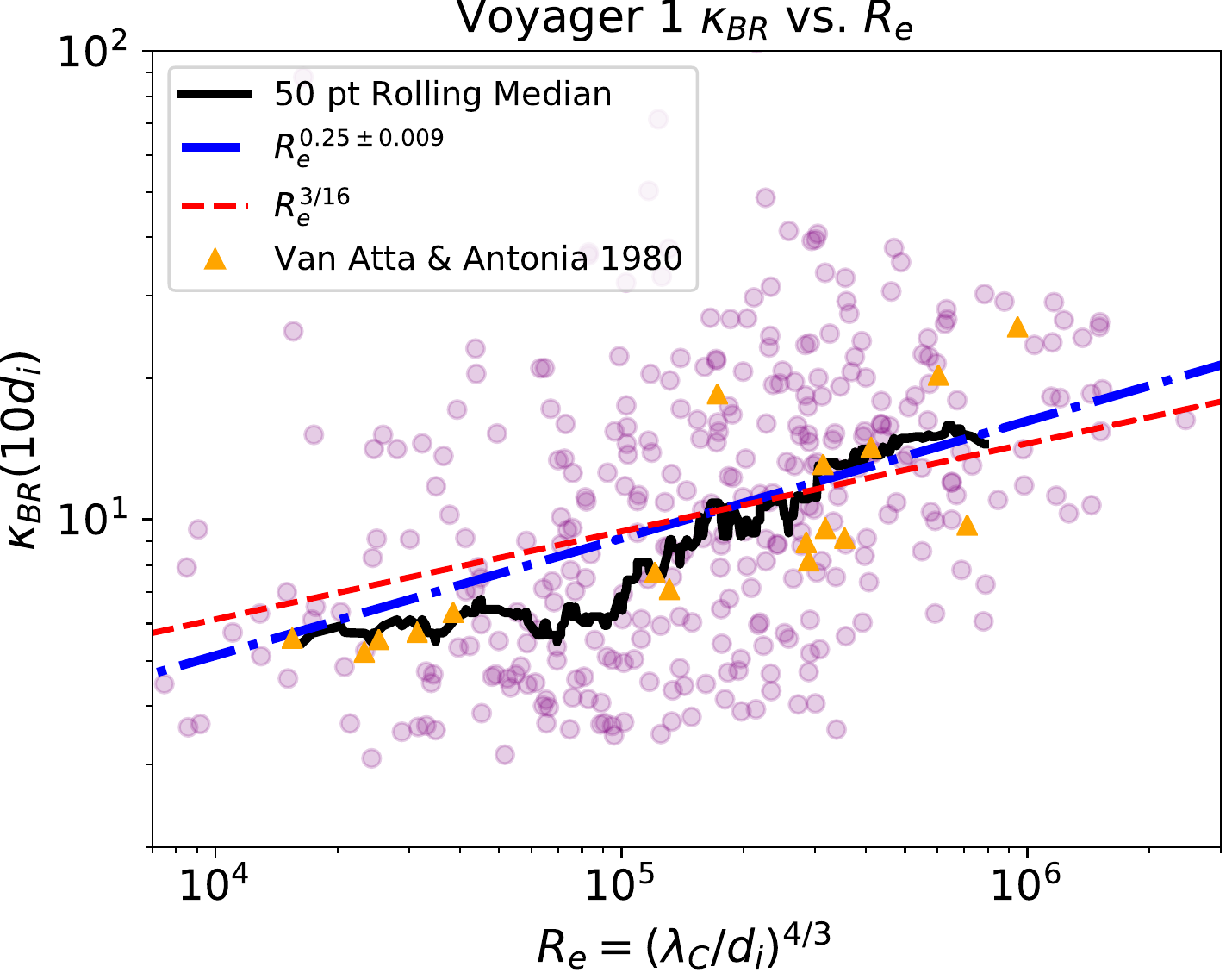}
        \caption{Kurtosis of radial magnetic field fluctuations as a function of ${\rm R_e}$ scaled at fixed lag $\ell=120~{\rm d_i}$ for Helios 1 (top) and $\ell=10~{\rm d_i}$ for Voyager 1 (bottom).  For comparison, orange triangles show 
        experimental hydrodynamic results 
        for velocity derivative kurtosis adapted from 
        \citet{VanAtta&Antonia1980} and re-scaled to the present (outer scale) Reynolds number dependence. The solid black line is a $50$ point rolling median and the red dashed line is an unfitted ${\rm  R_e}^{3/16}$ power-law.  The thicker blue dashed line is a fitted ${\rm  R_e}^{0.15}$ power-law with uncertainty $0.003$ for Helios 1 and a fitted ${\rm  R_e}^{0.25}$ power-law with uncertainty $0.009$ for Voyager 1. Power-law fits are computed over the range $10^5 < {\rm R_e} < 10^7$ for Helios and $10^4 < {\rm R_e} < 10^6$ for Voyager.}
        \label{fig:kbrvRe}
    \end{figure}

    \added{
    %{\bf 
    Let us once again assume that the radial dependence is $\kappa(\ell/{\rm d_i}) \sim R^\gamma$.
    We recall that ${\rm  R_e} \sim R^{-2/3}$ is reasonably well established on average, with however very large fluctuations,
    in both Helios and Voyager data.
    Therefore, due to the influence of these fluctuations we cannot conclude that $\kappa(\ell/{\rm d_i}) \sim {\rm R_e}^{-3\gamma/2}$.
    
    However, we can proceed to determine empirically the behavior of $\kappa_{BR}$ with fixed $\ell /{\rm d_i}$ for varying effective Reynolds number ${\rm  R_e}$, with no additional assumptions (e.g., log-normality).
    %}
    }
    The panels of Figure \ref{fig:kbrvRe} show the result of plotting the sampled observations of $\kappa$ vs ${\rm  R_e}$. 
    This is done for Helios observations (top) and Voyager (bottom).
    While there is considerable scatter, one may derive an average relationship from power-law fits (to these log-log scale distributions). 
    The result is $\kappa (\ell = 120 {\rm d_i}) \sim {\rm  R_e}^{0.15}$ for the Helios data, and $\kappa (\ell = 10 {\rm d_i}) \sim {\rm  R_e}^{0.25}$ for the Voyager data, with uncertainty $0.003$ and $0.009$, respectively.

    These determinations of kurtosis at relatively small inertial range scales are relevant to understanding intermittency and coherent structures, and their effects on the heating and dynamics of space plasmas.
    There are, however, as far as we are aware, 
    no analogous experimental  results in collisionless plasmas to which we may compare these observations.
    However, recent kinetic PIC simulation results
    \citep{ParasharEA2015}
    address this issue directly, although 
    the much smaller systems employed correspond to much lower effective Reynolds numbers.
    In particular, evaluating $\kappa(\ell = {\rm d_i})$
    for varying ${\rm  R_e}$, \citet{ParasharEA2015} found
    that $\kappa({\rm d_i}) \sim {\rm  R_e}^{a}$ with $a$ varying in a range of $\sim 0.1$ to $\sim 0.5$, 
    are computed in the range of ${\rm  R_e}$ from 10 to 100.
    The variation of SDK with ${\rm R_e}$ shown in Figure \ref{fig:kbrvRe} encompasses the lower end of this range, 
    even though 
    the range of ${\rm  R_e}$ in the spacecraft observations is at far higher values, between 
    $\sim 10^4$ to $\sim 10^6$.
    
    It is also of interest to adopt a wider turbulence perspective and compare the present results with the kurtosis of {\it longitudinal velocity derivatives} obtained in hydrodynamics experiments.
    A relevant hydrodynamics experimental result is $\kappa \sim {\rm R_{\lambda}}^{3/8}$ \citep{VanAtta&Antonia1980,Pope2000}, where ${\rm R_{\lambda}}$ is the Taylor-scale Reynolds number and $\kappa$ is based on measurements of the velocity derivatives component along the lag, $\partial u_\ell/\partial \ell$.
    Using a standard relation between Reynolds numbers at outer scale ${\rm R_e}$ and at Taylor scale, ${\rm R_{\lambda}}$, namely ${\rm R_{\lambda}}^2 \sim {\rm R_e}$, we arrive at a corresponding scaling with ${\rm R_e}$, i.e., $\kappa \sim {\rm  R_e}^{3/16}$.
    
    As mentioned above, there are no equivalent studies of ${\rm  R_e}$ variation of intermittency for plasmas. 
    Hence, here we will blur the distinction between $\kappa$ at small scale longitudinal increments and $\kappa$ for longitudinal derivatives, motivated by the finite time resolution of the Helios and Voyager spacecraft data.
    Also, the hydrodynamics result is for the velocity whereas we examine only the magnetic field kurtosis.
    With these caveats in mind, in Figure \ref{fig:kbrvRe}, we compare the hydrodynamics experimental measurements from \citet{VanAtta&Antonia1980} to both Helios and Voyager observations, which seem to be quantitatively consistent.
    \deleted{
    %{\bf 
    From Figure \ref{fig:kbrvRe}, both Voyager and Helios observed a weaker dependence of $\kappa_{BR}$ on ${\rm R_e}$ than found in the hydrodynamics system.
    %}
    }
    The average of the two fitted power-laws between Helios and Voyager lead to $\kappa_{BR} \sim {\rm R_e}^{0.2}$, compared to that of ${\rm R_e}^{3/16}$ expected from the hydrodynamics test case.
    For details regarding $\kappa_{BT}$ and $\kappa_{BN}$ at $\ell=120~{\rm d_i}$ for Helios and $\ell=10~{\rm d_i}$ for Voyager, as well as all $\kappa$ components at $\ell=120~{\rm d_i}$ for Voyager, see Figures \ref{fig:hl1_KvRe} and \ref{fig:vy1_KvRe} in Appendix \ref{app:SDK}.
    These additional results further demonstrate the similarities between MHD and log-normal hydrodynamics variation of $\kappa$ with ${\rm R_e}$. 
 
\subsection{Radial Variation of Intermittency Parameters $\mu(2)$ and $\mu(4)$}
\label{sec:mu}

    The observations described above can be used to evaluate the intermittency parameters $\mu(2)$ and $\mu(4)$ in several different ways. 
    A direct approach is to examine (fit) power-law behavior for both SF$_2$ and SF$_4$ from observations.
    Such results compare with expected behaviors for SF$_2$ and SF$_4$ from refined similarity, Equation \eqref{eq:magKRSH},  namely, $\ell^{\left( 2/3+\mu(2) \right)}$ and $ \ell^{\left( 4/3+\mu(4) \right)}$, respectively. 
    In Appendix \ref{app:SF}, Figures \ref{fig:SF2} and \ref{fig:SF4} show traces of SF$_2$ and SF$_4$ computed from 15 nearby useful intervals of our collection of Helios and Voyager observations.   
    This provides an appropriate ensemble average to empirically account for the associated second and fourth order intermittency parameters at varying heliocentric distance, by carrying out power-law fits in lag.
    We examine power-law fits in lag $\ell$ over \added{
    %{\bf 
    ${10-800}~{\rm d_i}$ for PSP,
    %}
    }
    ${10^3-10^4}~{\rm d_i}$ for Helios 1, and ${10^2 - 10^3}~{\rm d_i}$ for Voyager 1.
    These lag ranges are selected to lie within the inertial range.
    
    For convenience, unless otherwise specified, we let $\mu(n)$ indicate the intermittency parameters calculated from the separate power-law fits to SF$_2$ and SF$_4$ using solar wind observations.  
    The values obtained from observations may be compared with the expected values of the intermittency parameters obtained 
    from a log-normal theory, or an MHD adaptation of a log-Poisson theory, designated here as $\mu_{\rm ln}(n)$    
    and $\mu_{\rm MHD}(n)$, respectively.
    
\begin{figure}[htp!]
        \centering
        \includegraphics[scale=.5]{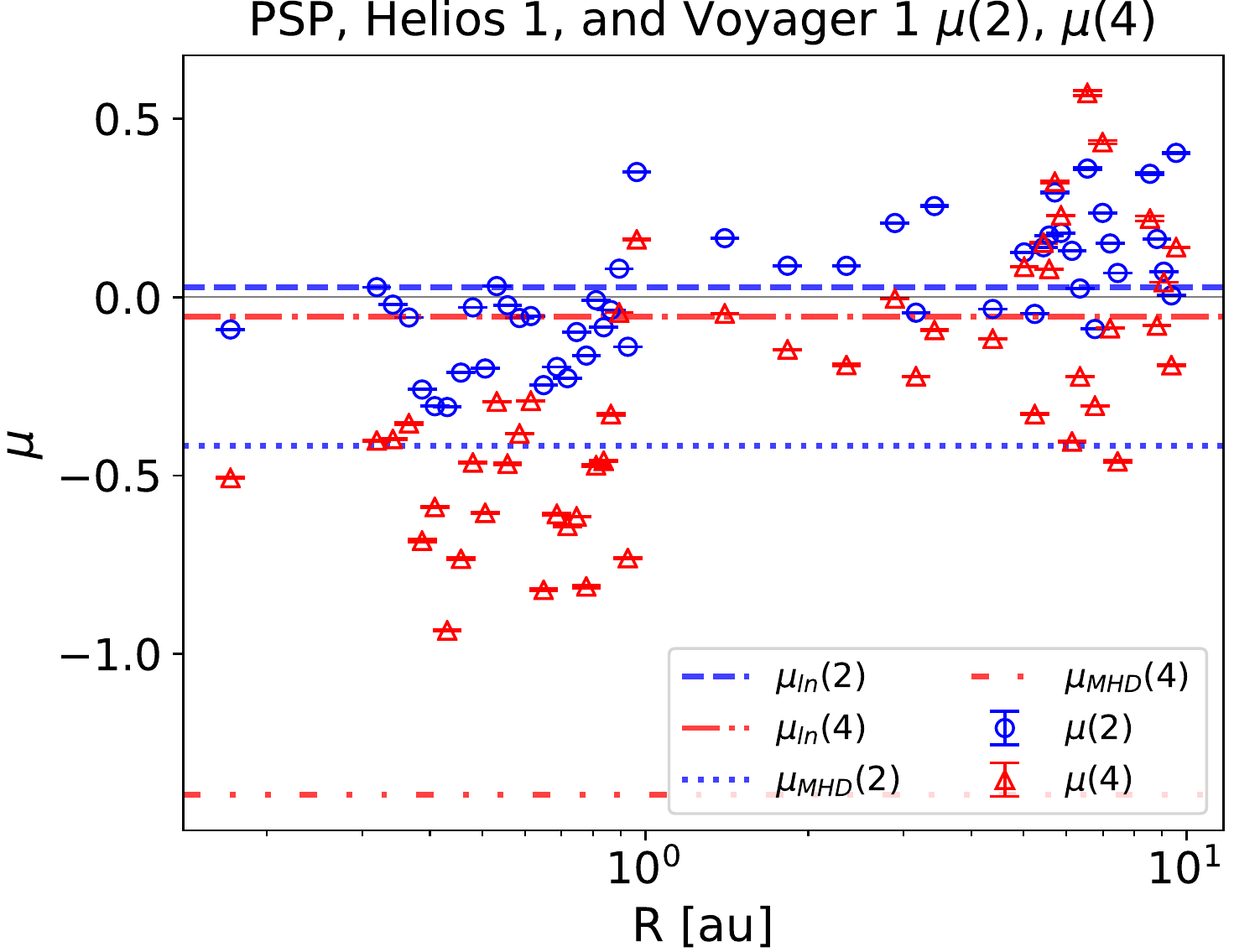}
        \vskip0.2in
        \includegraphics[scale=.5]{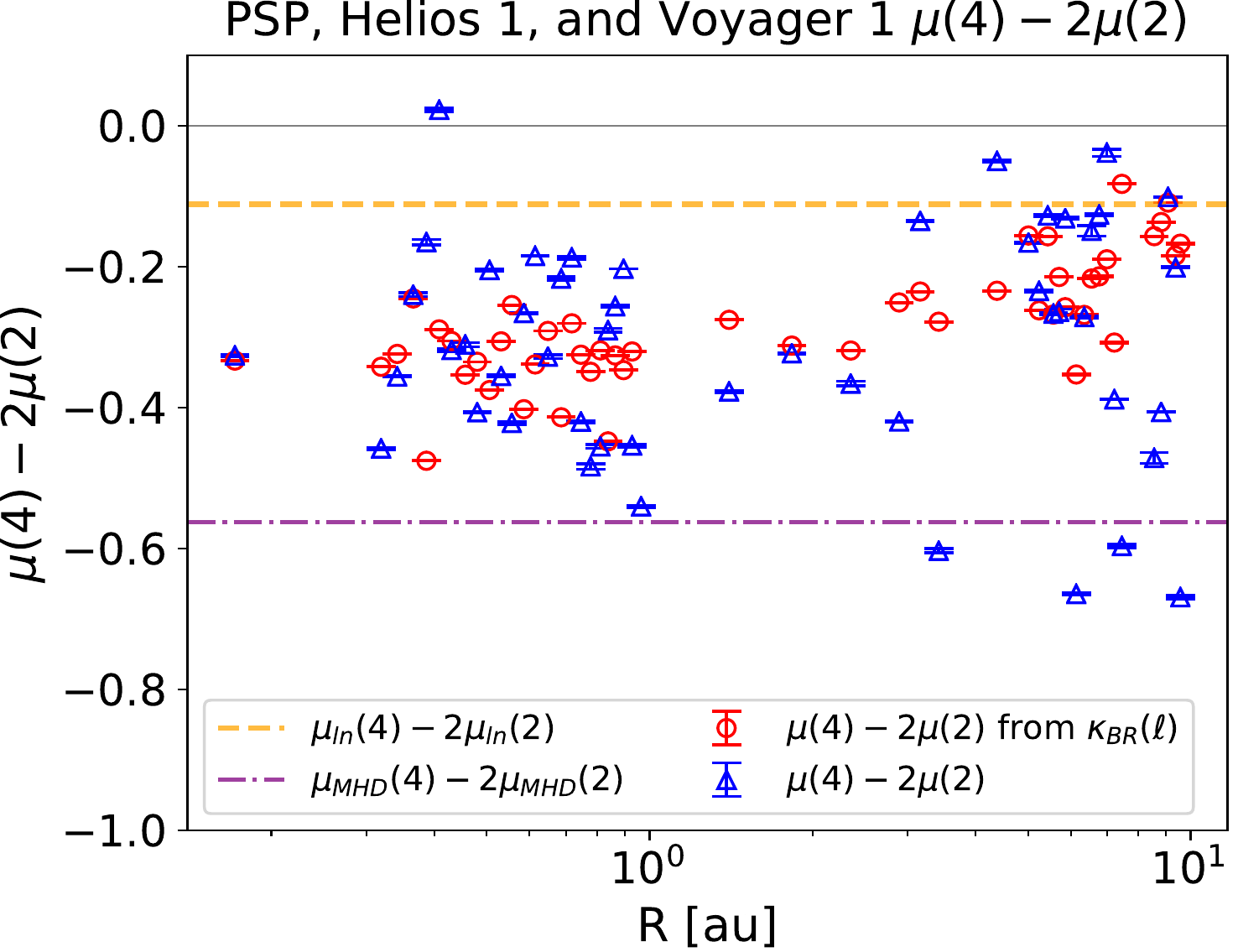}
        \caption { Heliocentric distance dependence of (top)
        $\mu(2)$ and  $\mu(4)$, and (bottom) $\mu(4)-2\mu(2)$. The data points are computed from fits over a range of
        lags spanning $10<\ell<800~{\rm d_i}$ for PSP, $10^3~{\rm d_i}<\ell<10^4~{\rm d_i}$ for Helios 1, and $10^2~{\rm d_i}<\ell<10^3~{\rm d_i}$ for Voyager 1. Values of $\mu (n)$ are computed from power-law fittings to SF$_n(\ell)$ from solar wind data. The exception is 
        the value computed from $\kappa_{BR}(\ell)$, which comes from direct fits to the observed kurtosis. The values  
        computed from theories (see text) are   $\mu_{\rm ln}(n)=\mu_0 n (3-n)/18$ for $\mu_0 \approx 0.25$ and 
        $\mu_{\rm MHD}(n)=1 - \frac{7}{12}n - 2^{-n}$. Error bars are given by the corresponding uncertainty for the fitted parameters.}
        \label{fig:mu}
    \end{figure}

    In the log-normal case, $\mu(n)= \mu_{\rm ln} (n) \equiv \mu_0 n (3-n)/18$ where we use the \citet{VanAtta&Antonia1980} value  $\mu_0 \approx 0.25$, and we let $\mu_{\rm ln}(n)$ indicate these values.
    \added{
    %{\bf 
    We also considered the log-Poisson model for comparison.
    This model is represented by \citep{She&Leveque1994}:
    
    \begin{equation} \label{eq:mulp}
        \mu_{\rm lp}(n)=-\frac{2}{9}n+2\left[ 1 - \left( \frac{2}{3} \right)^{n/3} \right].
    \end{equation}
    However, these log-Poisson and log-normal models were both derived to describe hydrodynamic fluids, which resulted in the values of $\mu_{\rm lp}(n)$ being nearly indistinguishable from $\mu_{\rm ln}(n)$ for $n<5$.
    Accordingly, we do not show the log-Poisson results in the figures.
    
    To provide an additional model to make a meaningful comparison with the log-normal model and the solar wind observations, we consider an MHD-intermittency model described in \citet{Marsch&Tu1997}.
    This model considers the \citet{She&Leveque1994} log-Poisson model with an adaptation for MHD that includes Kraichnan-Iroshnikov phenomenology.
    From this one can deduce an expression for the intermittency correction in parallel with $\mu_{\rm ln}(n)$, such that:
    
    \begin{equation} \label{eq:mu_MHD}
        \mu_{MHD}(n) = 1 - \frac{7}{12}n - 2^{-n}.
    \end{equation}
    This MHD model will serve as a suitable comparison to spacecraft observations given the similar MHD characteristics of the solar wind.
    We also directly examine the behavior of SDK with $\ell$, such that $\kappa \sim \ell^{\left( \mu(4) - 2\mu(2) \right)}$.
    The comparisons between model and observations are given in Fig. \ref{fig:mu}.
    %}
    }
  
    The top panel of Figure \ref{fig:mu} shows the results obtained for $\mu(2)$ and $\mu(4)$ computed from power-law fits to the observed SF$_n$, as a function of heliocentric distance $R$.
    For comparison, the approximate log-normal values $\mu_{\rm ln}(n)$ and $\mu_{\rm MHD}(n)$ are also exhibited. 
    Although it is not necessary for the solar wind and log-normal hydrodynamics values to be similar, $\mu_{\rm ln}(n)$ and $\mu(n)$ have similar values in some ranges.
    \added{
    %{\bf 
    The majority of values of $\mu(2)$ and $\mu(4)$ appear to lie between the log-normal and MHD scaling models for $R<1~{\rm au}$, whereas for larger radial distances, $\mu(2)$ and $\mu(4)$ are closer to the log-normal scaling model.
    %}
    }
    
    Next, we examine the variation of $\kappa$ with length scale, which is controlled by the combination of intermittency parameters $\mu(4) - 2\mu(2)$, a quantity that we will refer to as the {\it fractal scaling} of $\kappa$ (see \citet{ChhiberEA2021April}).
    We examine this scaling using four methods. 
    The first method is to compute power-law fittings to the measured kurtosis of the radial component of magnetic field $\kappa_{BR}(\ell/{\rm d_i})$.
    These fittings were computed over the aforementioned scales for PSP, Helios, and Voyager.
    The second, third, and fourth methods involve computing the fractal scaling using $\mu(n)$, $\mu_{\rm ln}(n)$, and $\mu_{\rm MHD}(n)$, respectively.
    Using Equations \eqref{eq:muln} and \eqref{eq:mu_MHD}, the third and fourth methods yield $\mu_{\rm ln}(4) - 2 \mu_{\rm ln}(2)\approx -0.\overline{11}$ and $\mu_{\rm MHD}(4) - 2 \mu_{\rm MHD}(2)\approx -0.56$.
    
    The bottom panel of Figure \ref{fig:mu} illustrates the fractal scaling of $\kappa$ from the spacecraft analyses. In nearly all solar wind regions $0.30~{\rm au} < R < 10~{\rm au}$, this scaling is larger in magnitude (more intermittent) compared to the log-normal hydrodynamics expectation, with one exception from using the second method.
    Also, nearly all fractal scalings lie between the log-normal and MHD models, with a trend towards log-normal scaling for increasing distance $R>1~{\rm au}$.
    \deleted{
    %{\bf 
    These exceptions occur at $R \approx 0.40~{\rm au}$ and  $R > 4~{\rm au}$.
    %}
    }
    This suggests that solar wind magnetic turbulence
    for $0.30~{\rm au} < R < 10~{\rm au}$ is generally more intermittent than log-normal
    hydrodynamics intermittency, but less intermittent than purely MHD intrmittency.
    Additionally, fractal scalings computed using the first method generally agree with those computed from the second method, with a few exceptions.
    For both decreasing heliocentric distance $R<1~{\rm au}$, and increasing distance $R > 1~{\rm au}$, there is a trend towards the log-normal hydrodynamics fractal scaling, although this trend is more evident for $R>1~{\rm au}$.
    In this sense, intermittency is getting stronger in the young solar wind, reaching a peak at distance $1~{\rm au} < R < 2~{\rm au}$, after which it begins to weaken for increasing distance (see discussion in Section \ref{sec:conclusions}).
    There are also few regions that suggest nearly mono-fractal scaling ($\mu(4) - 2 \mu(2) = 0$).
    Further observations for $R<0.30~{\rm au}$ using PSP and Solar Orbiter is expected to further characterize and clarify the current suggestions, which will be discussed below.
          
 In Section \ref{sec:sdkdi}, $\gamma$ was introduced to describe the 
 empirical determination of the radial scaling of the SDK at a fixed physical scale. However under the assumption of refined similarity, as in Section \ref{sec:SDK}, we understand that 
$\gamma$ becomes a function of $\mu(2)$ and $\mu(4)$.
In particular $\gamma = \left[ \mu(4) - 2\mu(2) \right]/2$.
Therefore the above results may be interpreted     
not only as a determination of fractal scaling, 
but also as quantifying 
how intermittency at a specified physical length scale 
becomes stronger or weaker with varying heliocentric distance $R$.
Results regarding dependencies of $\kappa(\ell/{\rm d_i})$ on $R$ and ${\rm R_e}$ were presented in Sections \ref{sec:sdkdi} and \ref{sec:sdkRe}, respectively.
 
%%%%%%%%%%%%%%%%%%%%%%%%%%%%%%%%%%%%%%%%%%
%%%%%%%%%%%%%%%%%%%%%
\section{Conclusions} \label{sec:conclusions}
%%%%%%%%%%%%%%%%%%%%%
    The main purpose of this study is to determine how small-scale intermittency in the inertial range evolves as the solar wind expands from the inner heliosphere ($0.16~{\rm au}$) out to $\approx 10~{\rm au}$.
    Firstly, we compute an effective Reynolds number
    ${\rm R_e}$ to quantify the system size from a turbulence perspective.
    This required computing auto-correlation functions of the magnetic field fluctuations for each interval to find its associated correlation time.
    Using the Taylor hypothesis, we convert these correlation time scales to spatial scales using the solar wind speed averaged over the interval.
    Then we compute ion inertial lengths using proton densities from those intervals to compute ${\rm R_e}$ using 
    the definition of 
    an effective Reynolds number given in Equation \eqref{eq:Re}.
    
    Secondly, we compute SF$_2$ to determine the scales spanning the inertial range; see Figure \ref{fig:SF2} in Appendix \ref{app:SF}.
    From SF$_2$ and SF$_4$ we calculate intermittency parameters $\mu(2)$ and $\mu(4)$, since SF$_n \sim \ell^{n/3+\mu(n)}$ assuming that refined similarity is applicable to magnetic increments.
    Furthermore, we compute scale-dependent kurtosis of each fluctuating magnetic field component by evaluating 
    its fourth-order structure function normalized by the square of its second-order structure function (cf. Section \ref{sec:SDK}).
    Once again referring to the refined similarity, 
    we expect that the kurtosis should behave as $\kappa(\ell) \sim \ell^{\mu(4)-2\mu(2)}$.
    However, to determine how ${\rm R_e}$ affects the strength of intermittency in a given interval, we make use of Equation \eqref{eq:sdkvelocity}, which rearranges terms to bring out variation with ${\rm  R_e}$ and $\ell$, with the additional assumption of inertial range log-normal statistics for the magnetic field.
    This provides that 
    $\mu_{\rm ln}(n)=\mu_0 n(3-n)/18$ for inertial range intermittency statistics.
    The value of $\mu_0 \approx 0.25$ used in this study is taken from 
    velocity field statistics in hydrodynamics experimental measurements \citep{VanAtta&Antonia1980} given a lack of equivalent studies for plasmas. It is remarkable that the hydrodynamic scaling describes the intermittency of magnetic fluctuations in plasmas this well.
    \added{
    %{\bf 
    It is also significant that the solar wind's fractal scaling is well contained within the range of values between the predicted log-normal and MHD scaling models, the latter being a log-Poisson hydrodyanmic model adapted to MHD utilizing Kraichnan-Iroshnikov phenomenology.
    %}
    }
    
As a general expectation
based on observations,
we expect that $\lambda_C \sim \sqrt{R}$ and ${\rm d_i} \sim R$; see Figures \ref{fig:CL} and \ref{fig:Di}, respectively.
Then the expected behavior of ${\rm R_e}$ 
is that ${\rm R_e} \sim R^{-2/3}$; see Figure \ref{fig:Re}.
    This line of reasoning suggests that the bandwidth of the inertial range is decreasing, i.e. the range of scales 
    between the correlation scale and the edge of the inertial range near the dissipation range is decreasing with increasing $R$.
    This conclusion is consistent with that of \citet{ParasharEA2019}, 
    extending those results to
    span heliocentric distances $0.16~{\rm au} < R < 10~{\rm au}$.
    
A point that warrants some brief discussion is the apparent
and somewhat precipitous decrease in correlation scale 
approaching PSP distances $\sim 0.16~{\rm au}$ 
from larger heliocentric distance, as seen in 
Figure \ref{fig:CL}.
There are several possible approaches to explaining this.
It has been noted for example that there
is a disparity in the radial evolution of 
correlation lengths measured parallel to the mean magnetic field and perpendicular to it. 
Although at $1~{\rm au}$ these lengths are typically nearly equal, with the parallel length possibly a factor of two larger \citep{DassoEA2005}, 
this seems not to be the case closer to the sun.
\citet{RuizEA2011} found in the Helios dataset 
that the measured
parallel correlations increase more rapidly with 
increasing heliocentric distance than do the perpendicular measurements.
(These are single spacecraft observations so 
the two orientations are not measured simultaneously.)
Notably the parallel correlation scales closer to Helios perihelion are considerably smaller than the perpendicular scales.

A similar kinematic issue is relevant to the 
PSP observations.   
Once again, as a single spacecraft mission, PSP
does not observe parallel and perpendicular scales simultaneously, but relies on variation of the ambient field direction. Therefore the frequency at which different angles are obtained strongly influences the separation of 
parallel and perpendicular correlations. 
Approaching perihelion the local magnetic field direction is seen, as expected, to become more dominantly radial and therefore observed correlation statistics become increasingly controlled by parallel sampling. This has been noted 
\citep{ChhiberEA2021Dec,ZankEA2021PoP} 
as a likely reason that PSP correlation scales near perihelion are systematically smaller than might be expected based on 
extrapolation from $1~{\rm au}$ observations and general trends from Helios observations. 
A recent turbulence transport calculation
based on a ``slab'' (parallel) 
and ``2D'' (perpendicular) representation of turbulence is able to reproduce 
the correlation scales found in selected parallel intervals in 
PSP data near perihelion \citep{AdhikariEAApril2021} with appropriate choice of parameters. 
This model result does not address the underlying question as to what is the physical cause of the smaller parallel correlation lengths nor the values of perpendicular scales (which are not observed). 

    Another line of reasoning relates to the issue of ``aging'' of the turbulence
    \citep{MatthaeusEAJGR1998}, or the degree to which it has evolved since energy was injected into the plasma.
Some of this injection probably occurs closer to the sun near the photosphere, while additional energy may be injected due to reconnection in the corona \citep{FiskKasper2020} or above the Alfv\'en zone \citep{RuffoloEA2020}.
In any case it is plausible that 
near PSP perihelia we observe turbulence that 
is younger and not yet fully developed, being closer to the site of energy injection.
This implies that disparities in parallel and perpendicular scales, as well as their rates of change with
radius may be related to the location and nature of the 
energy injection. The \citet{RuizEA2011} study organized the observations according to turbulence age, further clarifying the stated effect.

Another main result of the present paper, the radial development of kurtosis and intermittency,  
is also consistent with this perspective, namely that the inner heliosphere is characterized by an approach of turbulence to a fully developed state \citep{TelloniEA2021}. 
It can be seen, for example in the bottom panel of Fig. \ref{fig:mu},
that the degree of intermittency actually increases in the Helios orbits, moving from $0.30~{\rm au}$ to $1~{\rm au}$. 
A similar increase of the strength of intermittency with increasing radius was reported by \citet{CarboneEA2004}, consistent with recent work by \citet{AlbertiEA2020}. Eventually, as 
turbulence becomes fully developed, 
effects related to turbulence onset become less important.
Subsequently, with further increase of heliocentric distance, as described in Section \ref{sec:results}, the intermittency decreases on average as the 
corresponding effective Reynolds number, or system size, also decreases.  
The suggestion is that in the outer heliosphere, e.g., 
beyond $1~{\rm au}$, 
the turbulence remains in a fully developed state during the gradual decrease in Reynolds number with increasing distance. 
Meanwhile at lower heliocentric distances, 
in the inner heliosphere, the evolution of correlation
scales may be due to 
onset of a
fully developed state, as well as sampling preferentially along the mean magnetic field. 
Of course neither of the latter explanations explain in any obvious way why the parallel correlation 
scales are relatively 
small compared to extrapolations from larger radii. 

The results found here for 
intermittency and scale dependent kurtosis 
(SDK) parallel the ideas discussed above 
regarding correlation scales, but 
perhaps are a bit more complex. 
The results in 
Section \ref{sec:SDK} indicate that 
    generally higher values of kurtosis 
    are found at smaller spatial lags.
Moreover, 
at a given scale, and on average, 
larger kurtosis values are found at smaller 
heliocentric distance, especially for $\kappa_{BR}$.
The magnitude of kurtosis is a measure of the non-Gaussianity,
or fat tails, on the the PDF of increments.     
So this diagnostic clearly 
supports familiar ideas that 
intermittency increases at smaller scales.
Less familiar is the strong suggestion that 
intermittency, measured by kurtosis,
decreases moving towards large heliocentric distances. 
We attribute this to a decreasing effective Reynolds number.

As PSP explores inner regions of the heliosphere for the first time, the potential emerges for developing a more detailed 
picture of the evolution of turbulence in the entire heliosphere.  
Such a global perspective may be useful in 
understanding how the solar wind is accelerated and heated, and 
also how turbulence evolves to the outer reaches of the sun's plasma environment.  
This perspective may be particularly valuable in understanding how energetic particles -- both the solar energetic particles and the galactic cosmic rays -- are scattered and transported throughout the volume of the heliosphere \citep{OughtonEA2021}. 
The present paper has made an attempt to take steps towards development of this more global perspective on the dynamics and evolution of turbulence, an effort that eventually may unify 
PSP observations of the inner heliosphere and 
emerging descriptions of the outer heliosphere obtained by 
Voyager, IBEX, and the upcoming IMAP missions.

%------------------------------------------------------
\section{Acknowledgements}
    This work is supported in part by the NASA Heliophysics Guest Investigator Programs (80NSSC19K0284 and 80NSSC21K1765), by NASA Supporting Research (80NSSC18K1648), by the 
    NASA PUNCH project (SWRI subcontract N99054DS), the LWS program (New Mexico Consortium subcontract 655-001), the 
    IS\(\odot\)IS Parker Solar Probe Project through Princeton subcontract SUB0000165, and the IMAP mission under Princeton subcontract
    SUB0000317.


\newpage
\newpage
\newpage
\appendix

%%%%%%%%%%%%%%%%%%%%%%%%%%%%%%%%%%%%%%%%%%
%%%%%%%%%%%%%%%%%%%%%%%%%%%%%%%%%%%%%%%%%%
\section{Data Specifics}\label{app:data}

    We employ magnetic field data from the PSP, Helios 1, and Voyager 1 spacecraft,
    obtained 
    from the Goddard Space Flight Center Space Physics Data Facility (SPDF).
    Time resolutions are $6.8~{\rm ms}$ for PSP, $6~{\rm s}$ for Helios 1, and $1.92~{\rm s}$ for Voyager 1.  These differences in available time cadences impact the details of the analysis described in the main text. 
    Other relevant parameters -- heliocentric distance, proton density, and proton speed -- are retrieved from SPDF at resolutions of $0.874~{\rm s}$ for PSP and $1~{\rm hr}$ for both Helios 1 and Voyager 1.
    The only utilization of these three plasma parameters entailed taking averages over an interval length to describe the interval's radial distance, ion inertial length, and factor of conversion between spatial and temporal via Taylor's hypothesis.
    
%%%%%%%%%%%%%%%%%%%%%%%%%%%%%%%%%%%%%%%%%%
\subsection{Parker Solar Probe}\label{subapp:PSP}

    Level 2 PSP magnetic field data from was extracted from SPDF 
    in RTN coordinates at full cadence.  
    We use measurements made by the fluxgate magnetometer onboard the FIELDS instrument suite \citep{BaleEA2016} and by the SPC onboard the SWEAP instrument \citep{KasperEA2016}.
    Three ${\rm 6 hr}$ intervals were downloaded, one from each of the first three PSP perihelia.  
    The intervals have start times
    "2018-11-06-00:00" for perihelion 1, "2019-04-04-18:00" for perihelion 2, and "2019-08-30-12:00" for perihelion 3.
    We resampled the data as needed 
    to the desired resolution of ${\rm 6.8 ms}$.
    Heliocentric distance and Level 3 proton density and speed data
    did not require resampling.
    
%%%%%%%%%%%%%%%%%%%%%%%%%%%%%%%%%%%%%%%%%%
\subsection{Helios 1}\label{subapp:HL1}

    Helios 1
    magnetic field data 
    was extracted from SPDF 
    in RTN coordinates and 
    at a $6~{\rm s}$ resolution for all available times from the years between 1974 to 1981.
    The measurements used in this study are taken from the magnetic field \citep{NeubauerEA1977} and plasma \citep{RosenbauerEA1977} instruments onboard Helios 1.
    Data gaps occurred throughout this dataset. Accordingly,
    we resampled to a regularly sampled series at the expected $6~{\rm s}$ resolution.  
    The gaps were filled with NaNs; sub-intervals with excessive NaNs were discarded (see Section \ref{subapp:comp}).  Resampling was not required for the heliocentric distance, proton density, and proton speed, and these 
    parameters were downloaded from SPDF at an hourly averaged resolution.
    \added{
    %{\bf 
    From a total of $2375$ intervals and $\approx 34.2~{\rm million}$ data points, $771$ intervals were useful and $\approx 28.3~{\rm million}$ points are NaN.
    %}
    }

%%%%%%%%%%%%%%%%%%%%%%%%%%%%%%%%%%%%%%%%%%
\subsection{Voyager 1} \label{subapp:VY1}
    The existing archival 
    Voyager 1 magnetic field data was extracted from SPDF in RTN coordinates
    and at a $1.92~{\rm s}$ resolution for times ranging heliocentric distances $1-10~{\rm au}$.
    A dual low-field and high-field magnetometer system was used for the Voyager mission \citep{BehannonEA1977}, as well as the plasma instrument \citep{BridgeEA1977}.
    Although missing data and data gaps were encountered, we discovered several other inaccuracies with respect to the quality of the data.  
    The distribution of consecutive time increments suggests the possibility that the spacecraft magnetometer system mislabeled the records in an irregular fashion.  
    To resolve these issues, we resampled the downloaded data to a regularly sampled time series with the expected $1.92~{\rm s}$ resolution.
    
    The next major issue encountered with the original RTN magnetic field data involve calibrations.  
    Unable to efficiently remove these via an autonomous process, manual effort was required in locating the intervals pertaining to these calibrations.  
    During the manual sifting process over the $\approx 60$ million data points, we also flagged intervals containing instrument noise and planetary encounters.
    The properties of the solar wind are different from that of near-planet regions, and therefore will interrupt radial trends.  
    As a result, calibrations, noise, and planetary encounters were removed.
    To finish improving the Voyager 1 magnetic field data, a type of Hampel filter \citep{Pearson2002} was applied.  
    Since all of the previously listed issues rendered the original dataset unusable, a Hampel-type filter was used to remove any other possible inaccuracies that may have remained undiscovered throughout the cleaning process.  
    The Hampel-type filter removed any statistical outliers beyond a $2.75$ multiple of the standard deviation from the mean within a $100$ point window.  
    This window sifted through the time series by $1$ point increments.  
    \replaced{\sout{The entire description of the cleaning process, including  justifications for choice of these parameters and cleaning methods will be found on SPDF once the refined data product is made public, or as requested \citep{MyMSThesis}.}}
    {
    %{\bf
    The entire description of the cleaning process, including justifications for choice of these parameters and cleaning methods, can be found in \citet{MyMSThesis}.  The dataset itself can be accessed at \href{https://doi.org/10.5281/zenodo.5711178}{https://doi.org/10.5281/zenodo.5711178}.
    %}
    }
    \added{
    %{\bf 
    From a total of $666$ intervals and $\approx 60~{\rm million}$ data points, $407$ intervals were useful and $\approx 36.7~{\rm million}$ points are NaN.
    %}
    }
    
%%%%%%%%%%%%%%%%%%%%%%%%%%%%%%%%%%%%%%%%%%
\subsection{Computation Specifics}\label{subapp:comp}

    \added{
    %{\bf
    For Helios 1 and Voyager 1,
    we divided the 
    data into 24 and 48 hour intervals, respectively.  For this study, we choose 24 and 48 hour interval sizes instead of the $0.02~{\rm au}$ radially binned interval size used in 
    \citet{ParasharEA2019}.
    %}
    }
    Although an interval's correlation time is systematically 
    affected by the interval's size \citep{IsaacsEA2015,JagarlamudiEA2019}, there is no significant change in the quantitative 
    results 
    compared to \citet{ParasharEA2019}, and the qualitative assessments of the observations remain 
    unchanged.
    Results for PSP data intervals
    described in Section \ref{subapp:PSP} made use of six hour intervals.   
    Additionally, there have been other efforts using PSP data resampled to a $1~{\rm s}$ resolution that involve computing correlation lengths with interval sizes as large as a day \citep{ParasharEA2020}.
    Averages of heliocentric distance, proton density, and proton speed over each of these intervals were used in computing spatial scales.
    
    In terms of computing the statistical quantities themselves, any interval composed of less than 20 percent non-NaN points were not included in the analysis.  
    This maintains a sufficient statistical weight within any given interval.  
    For those intervals that pass this prerequisite, there are specific limitations for computing intermittency statistics.  
    It is possible for the averages of the proton speed and proton density to yield NaN over any given interval.  
    If the average proton density is NaN, it is not possible to compute a value for ${\rm d_i}$, nor is it possible to apply the Taylor hypothesis if the average proton speed is NaN.
    As a result, we do not include 
    intervals with either of these 
    shortcomings
    in this analysis.

%%%%%%%%%%%%%%%%%%%%%%%%%%%%%%%%%%%%%%%%%%%%%%%%%%%%%%%
\newpage
\section{Supplementary Material for Structure Function}\label{app:SF}

\begin{figure*}[htp!]
        \centering
        \gridline{\fig{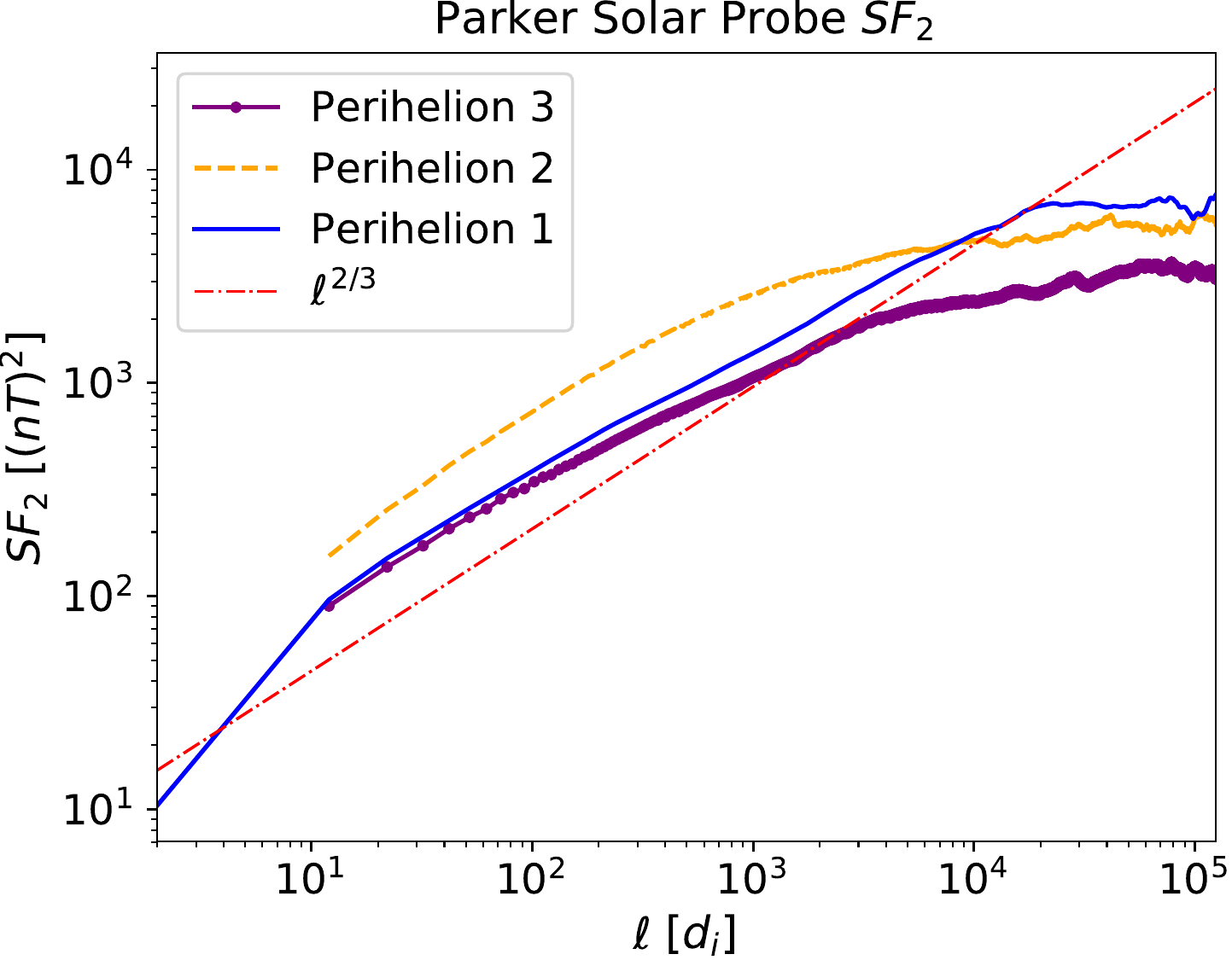}{0.3\textwidth}{(A)}
                \fig{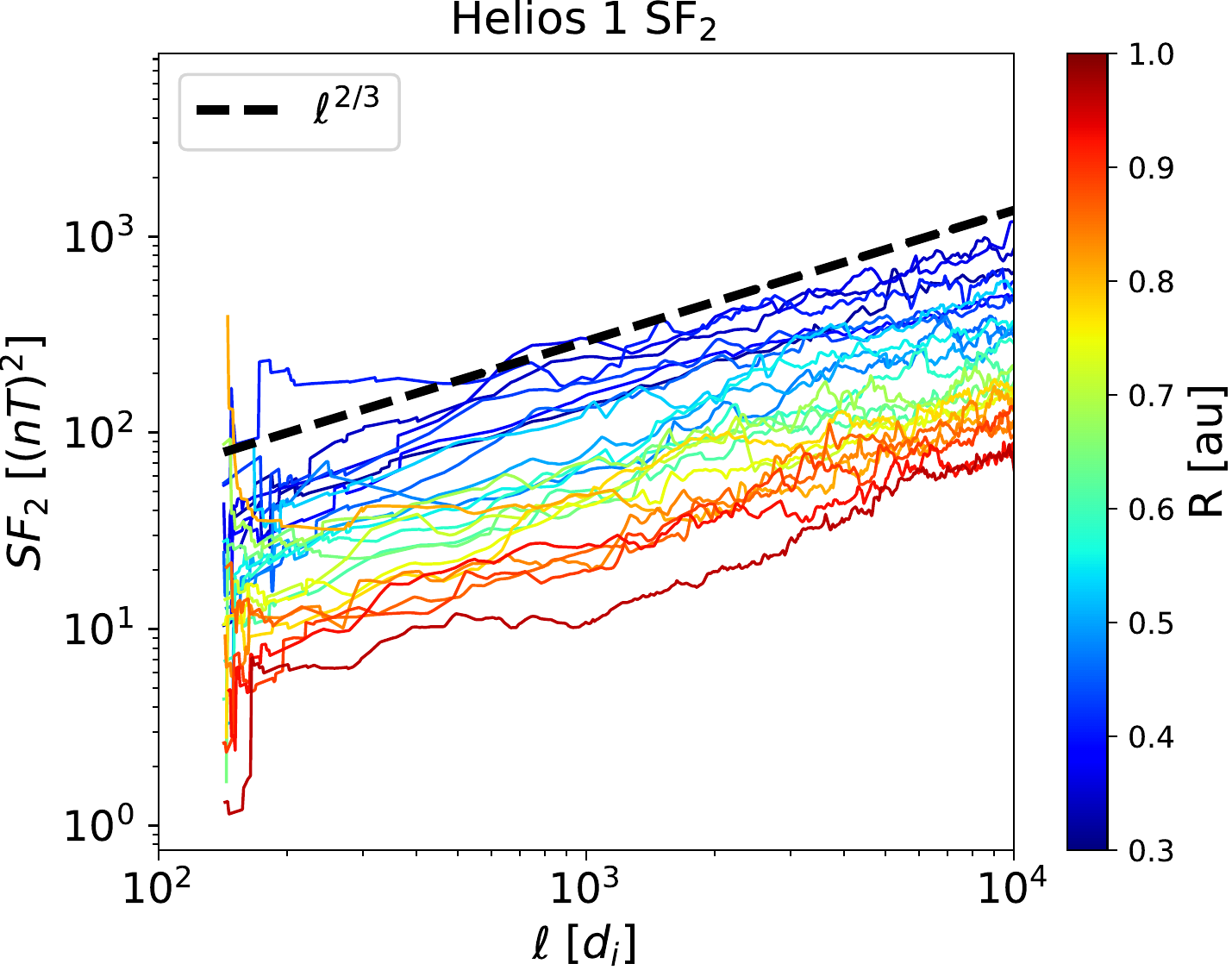}{0.3\textwidth}{(B)}
                \fig{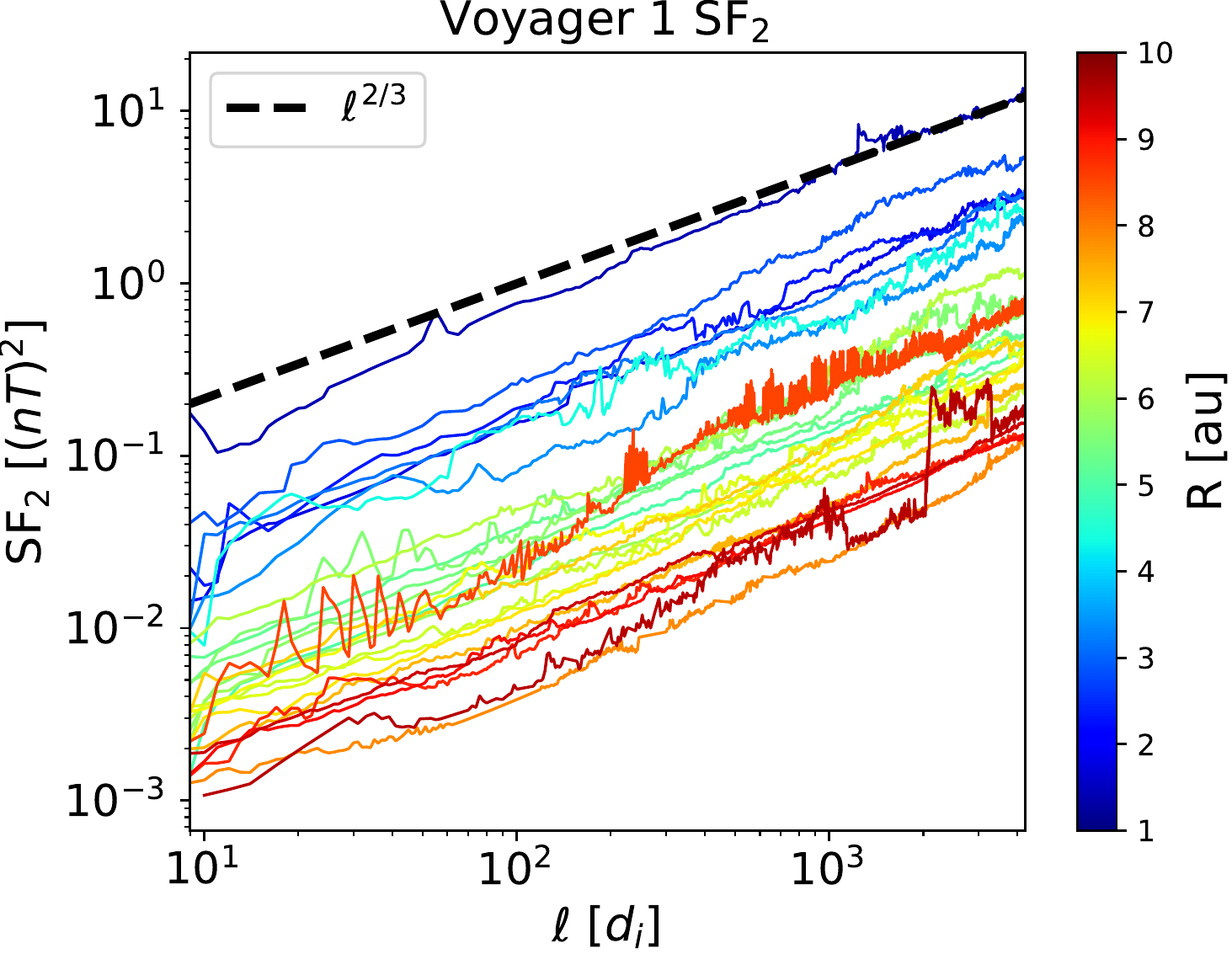}{0.3\textwidth}{(C)}
        }
        \caption{Second-order magnetic structure function for (A) PSP, (B) Helios 1, and (C) Voyager 1.}
        \label{fig:SF2}
    \end{figure*}

\begin{figure*}[htp!]
        \centering
        \gridline{\fig{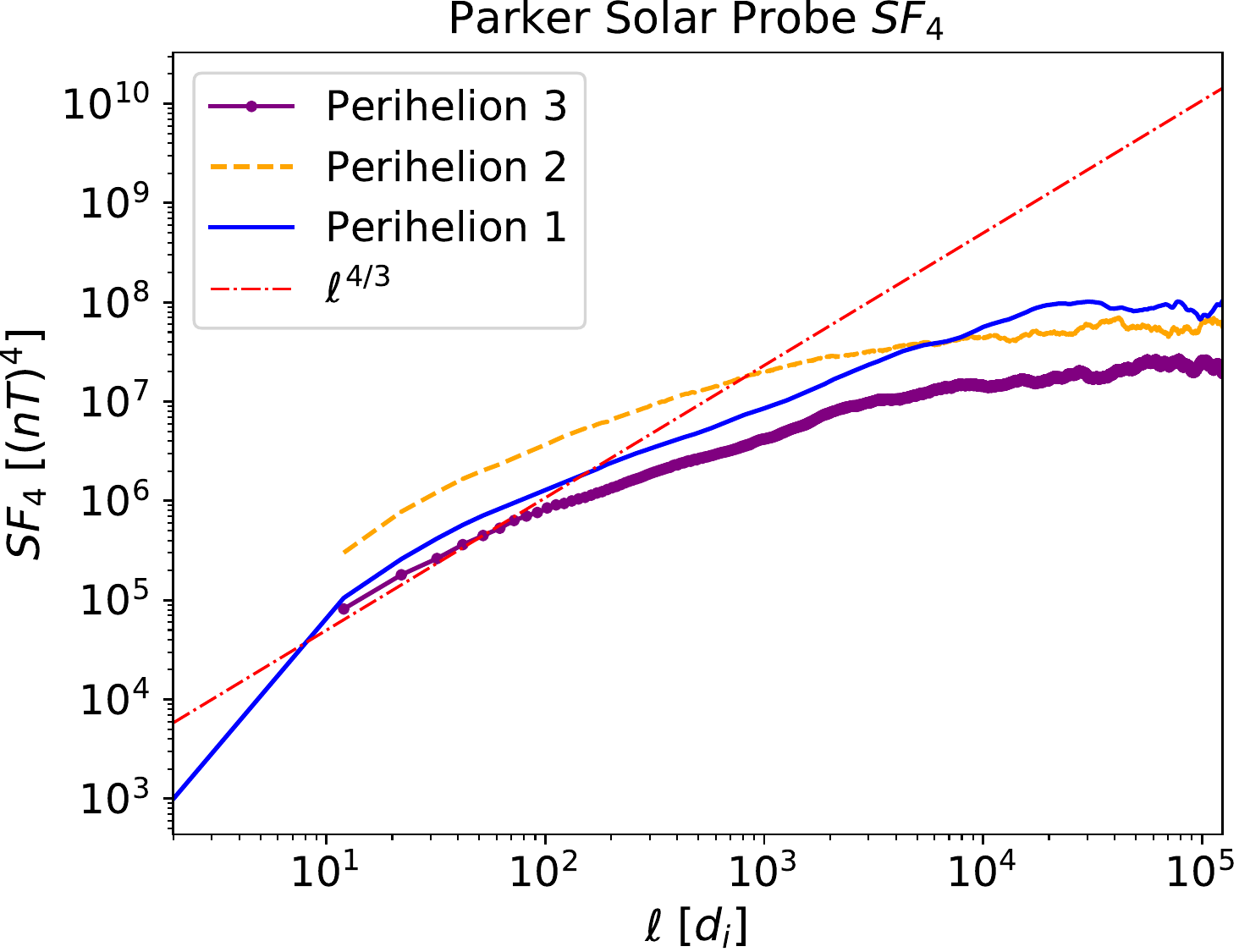}{0.3\textwidth}{(A)}
                \fig{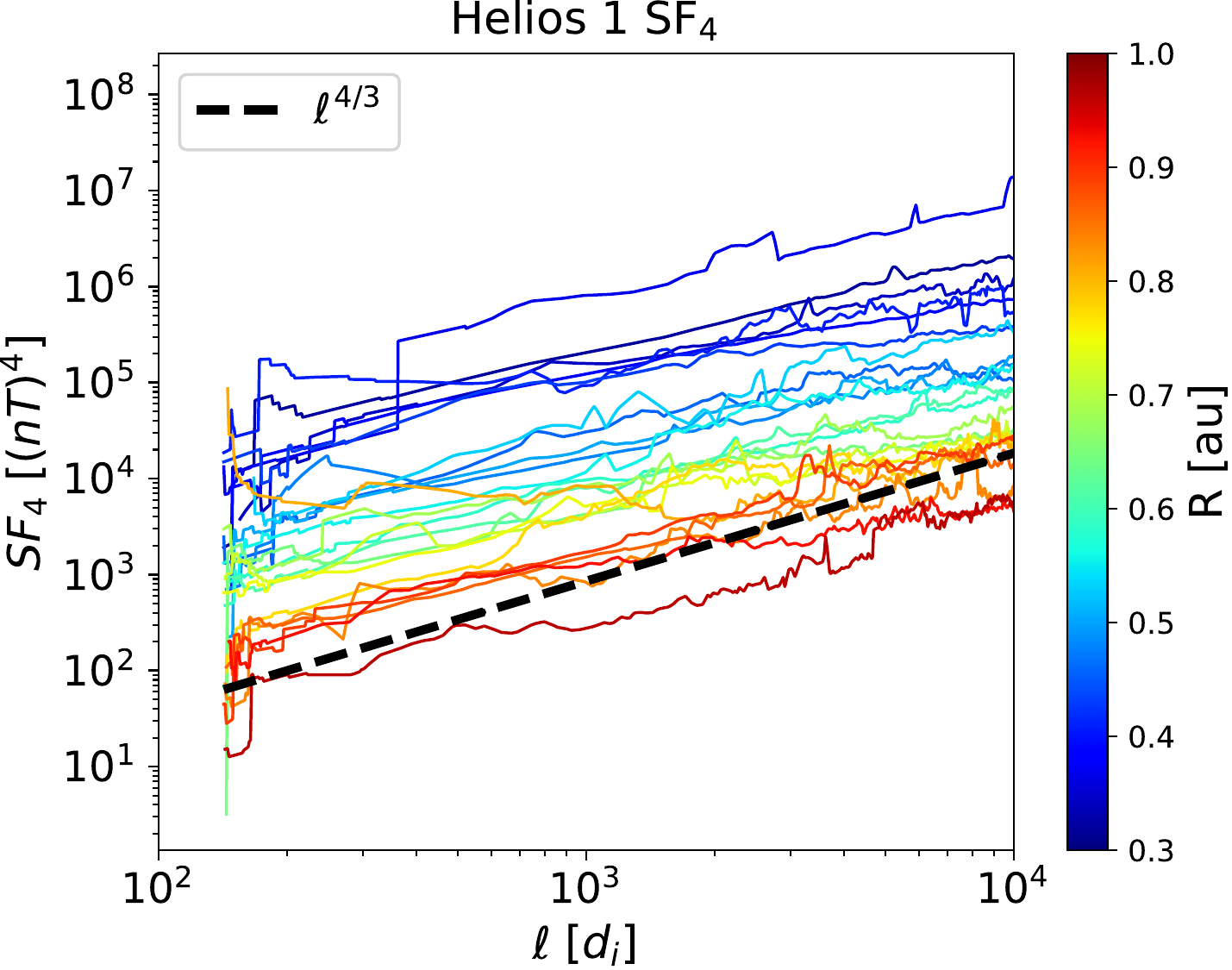}{0.3\textwidth}{(B)}
                \fig{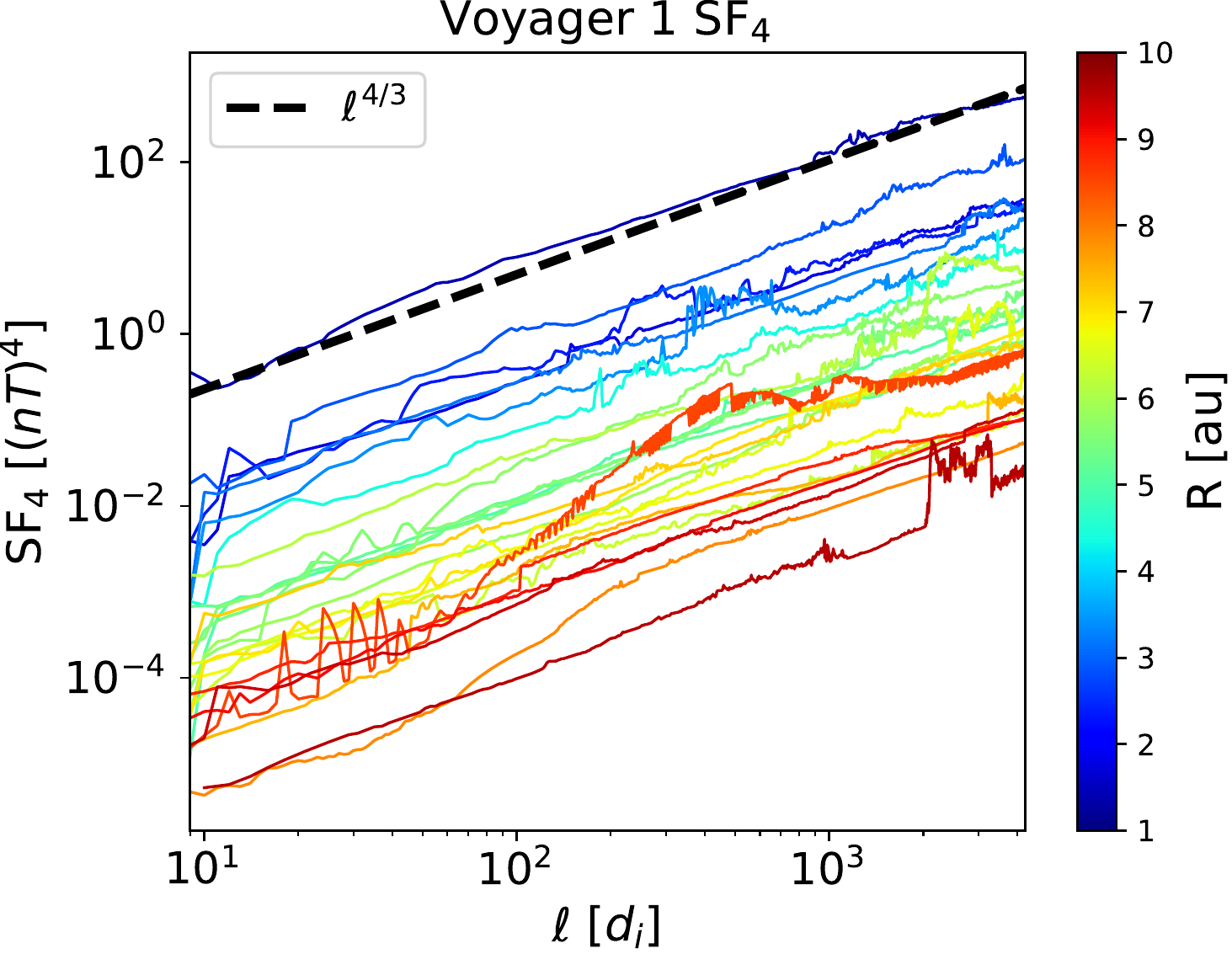}{0.3\textwidth}{(C)}
        }
        \caption{Fourth-order magnetic structure function for (A) PSP, (B) Helios 1, and (C) Voyager 1.}
        \label{fig:SF4}
    \end{figure*}

    Here we exhibit the magnetic structure functions for $n=2,4$ for PSP, Helios 1, and Voyager 1.
    The curves in Figures \ref{fig:SF2} and \ref{fig:SF4} for Helios 1 and Voyager 1 are averages of 15 nearby useful intervals, whose slopes are used to calculate intermittency parameters $\mu(2)$ and $\mu(4)$, respectively, as described in Section \ref{sec:mu}.
    These $\mu(2)$ and $\mu(4)$ values can be found in the top panel of Figure \ref{fig:mu}.
    For more details regarding SF$_n$, see Section \ref{subsec:SF}.
    Note that the power-law form of SF$_2$ is reasonably well defined for Helios and Voyager over a wide range of lag $\ell$.


%%%%%%%%%%%%%%%%%%%%%%%%%%%%%%%%%%%%%%%%%%%%%%%%%%%%%%%%%

\newpage
\section{Supplementary Material for Scale-Dependent Kurtosis and its Variation with Effective Reynolds Number at a Fixed Scale}\label{app:SDK}

\begin{figure*}[htp!]
    \centering
    \gridline{\fig{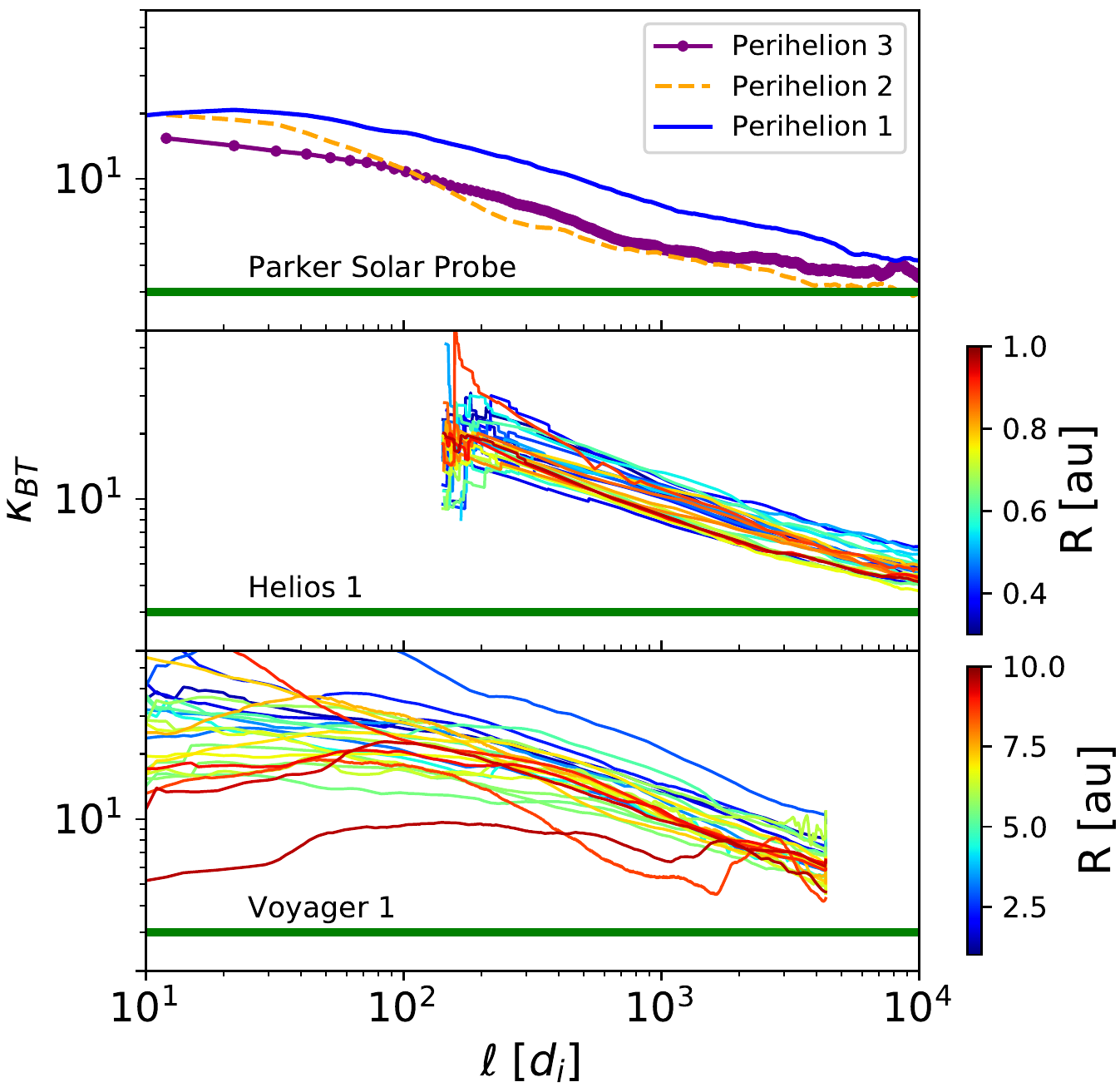}{0.45\textwidth}{(A)}
            \fig{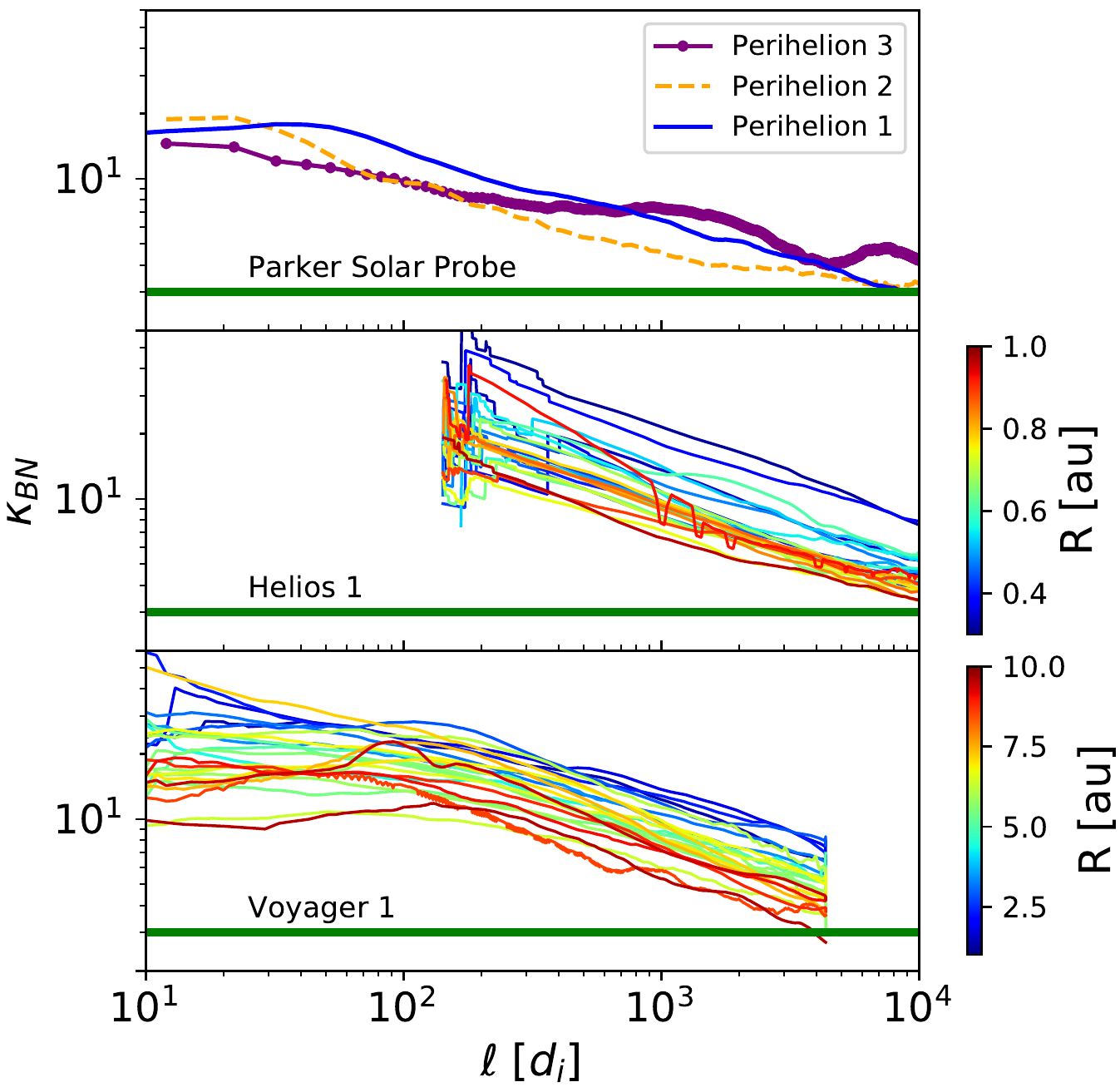}{0.45\textwidth}{(B)}
    }
    \caption{SDK of the tangential (A) and normal (B) magnetic field fluctuating component for PSP (top), Helios 1 (middle), and Voyager 1 (bottom). The color of the curve for Helios and Voyager are keyed to heliocentric distance.}
    \label{fig:SDKBTBN}
\end{figure*}

In addition to the radial magnetic component of SDK presented in Section \ref{sec:SDK}, we also provide the tangential and normal magnetic components for PSP, Helios 1, and Voyager 1 in Figure \ref{fig:SDKBTBN}.
Each curve in Figure \ref{fig:SDKBTBN} for Helios and Voyager are averages of 15 nearby useful intervals.
For more details regarding SDK, see Section \ref{subsec:SDK}.

%%%%%%%%%%%%%%%%%%%%%%%%%%%%%%%%%%%%%%%%%%%%%%%%%%%%%%%%%%%%%%

\begin{figure*}[htp!]
    \centering
    \gridline{\fig{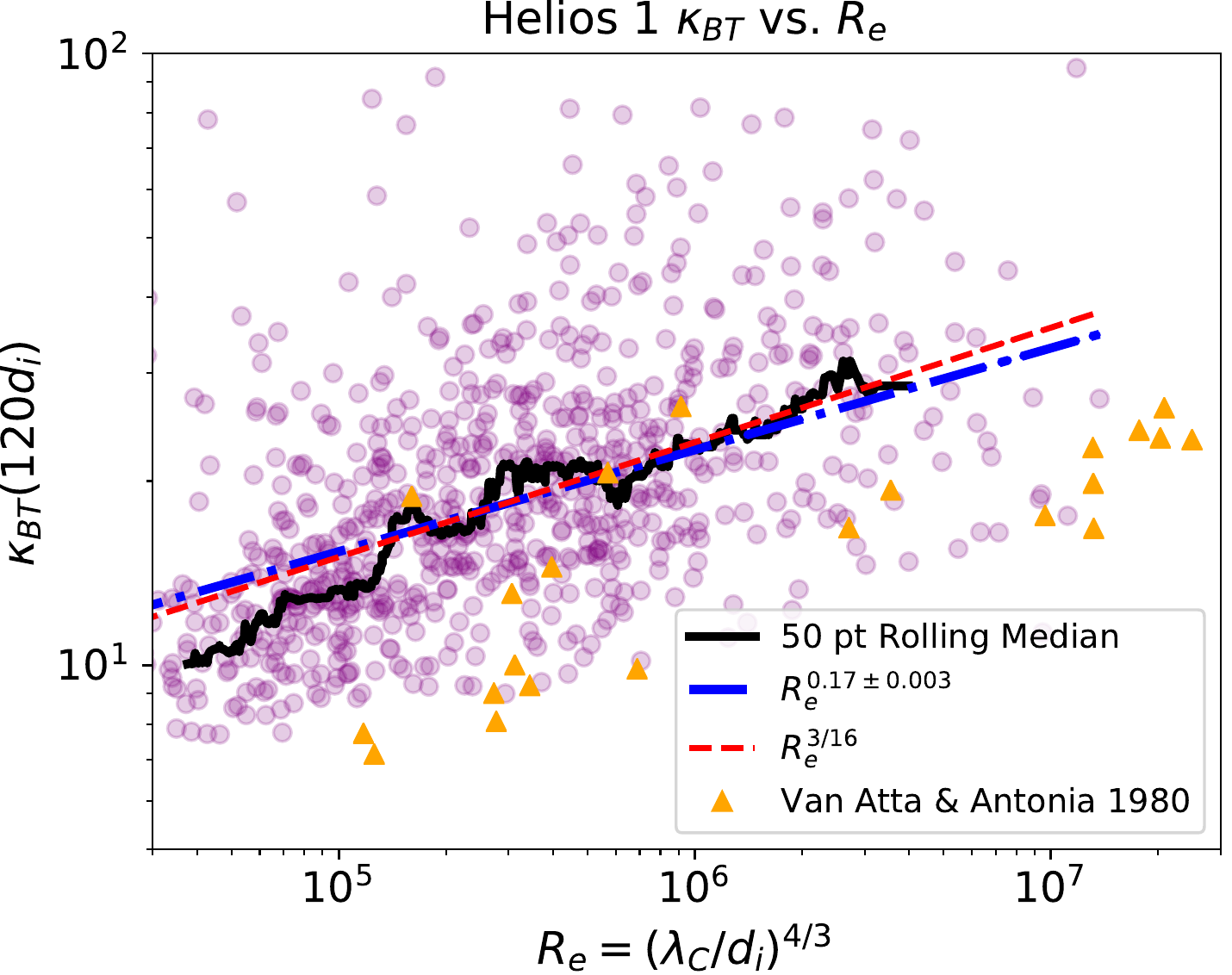}{0.3\textwidth}{(A)}
            \fig{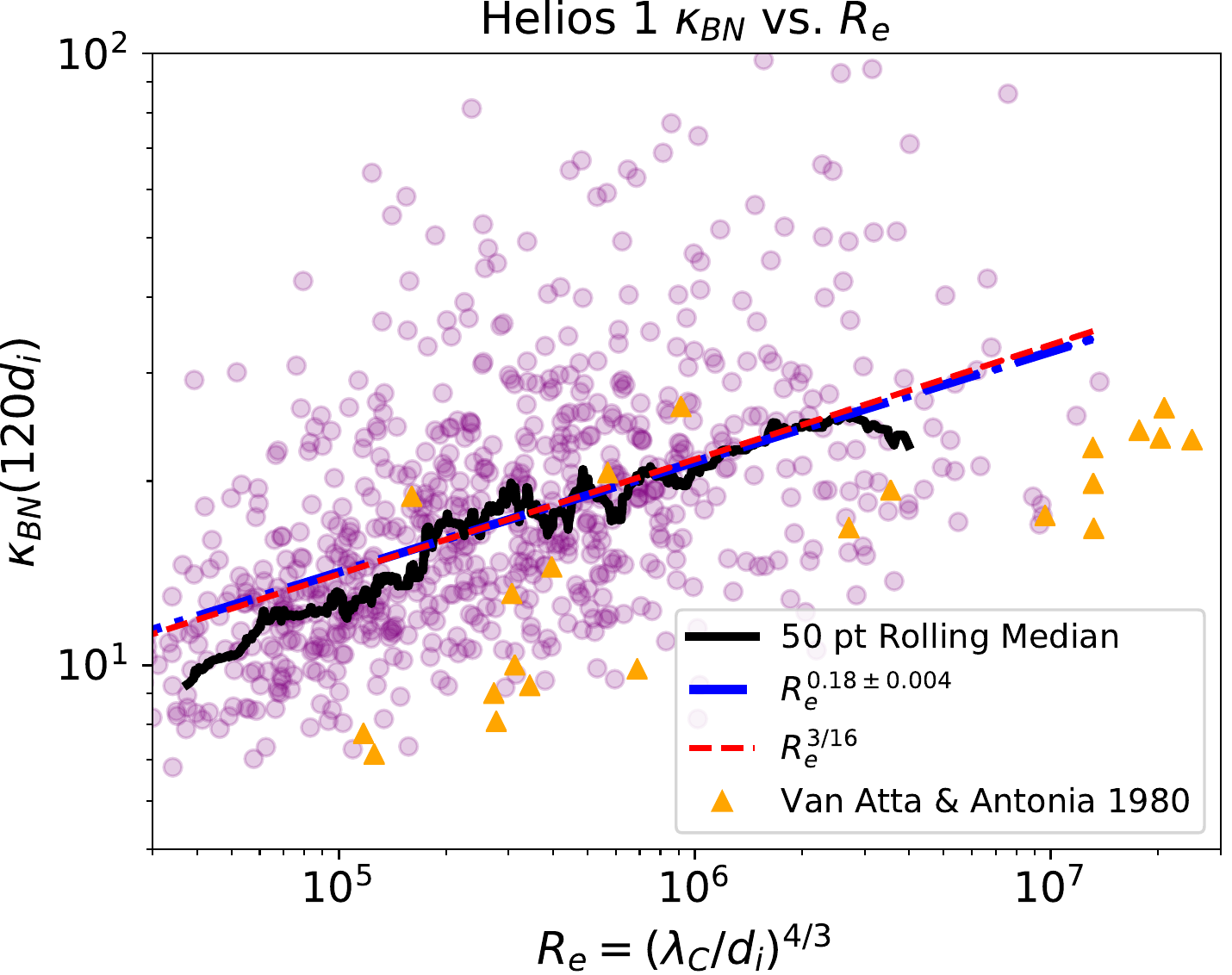}{0.3\textwidth}{(B)}
    }
    \caption{Kurtosis of the tangential (panel A) and normal (panel B) magnetic field fluctuating component held at $120~{\rm d_i}$, as a function of ${\rm  R_e}$, for Helios 1.}
    \label{fig:hl1_KvRe}
\end{figure*}

\begin{figure*}[htp!]
    \centering
    \gridline{\fig{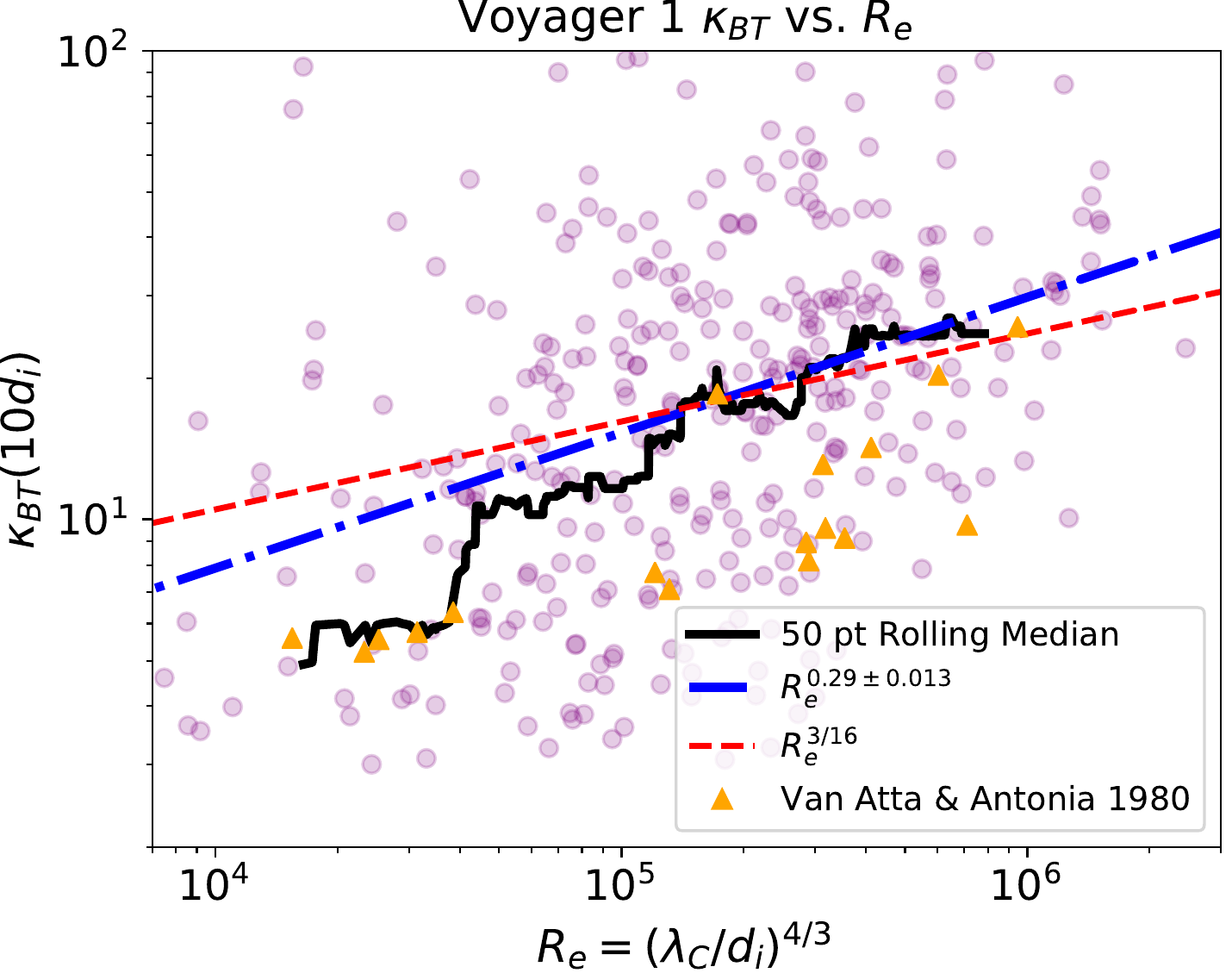}{0.3\textwidth}{(A)}
            \fig{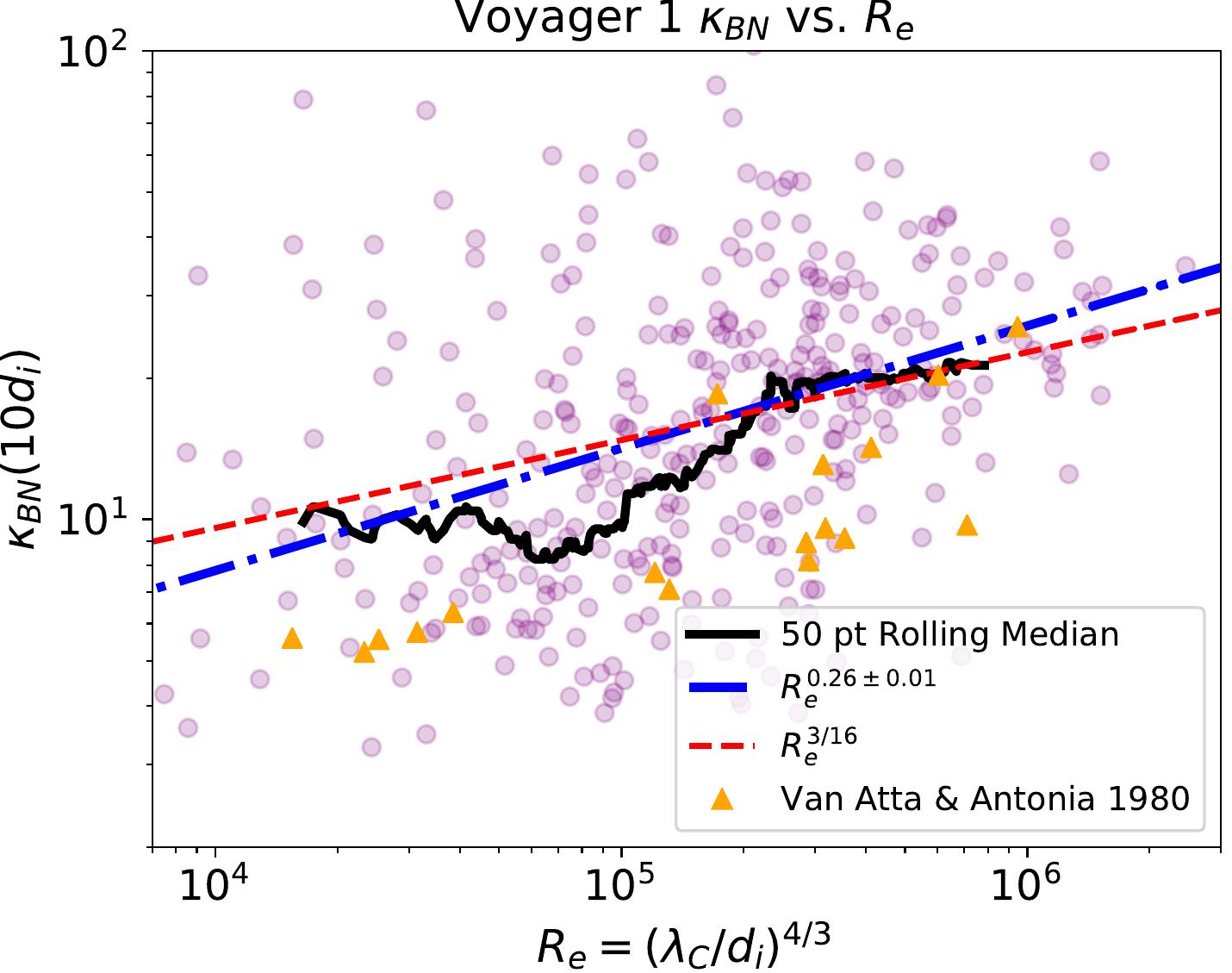}{0.3\textwidth}{(B)}
    }
    \gridline{\fig{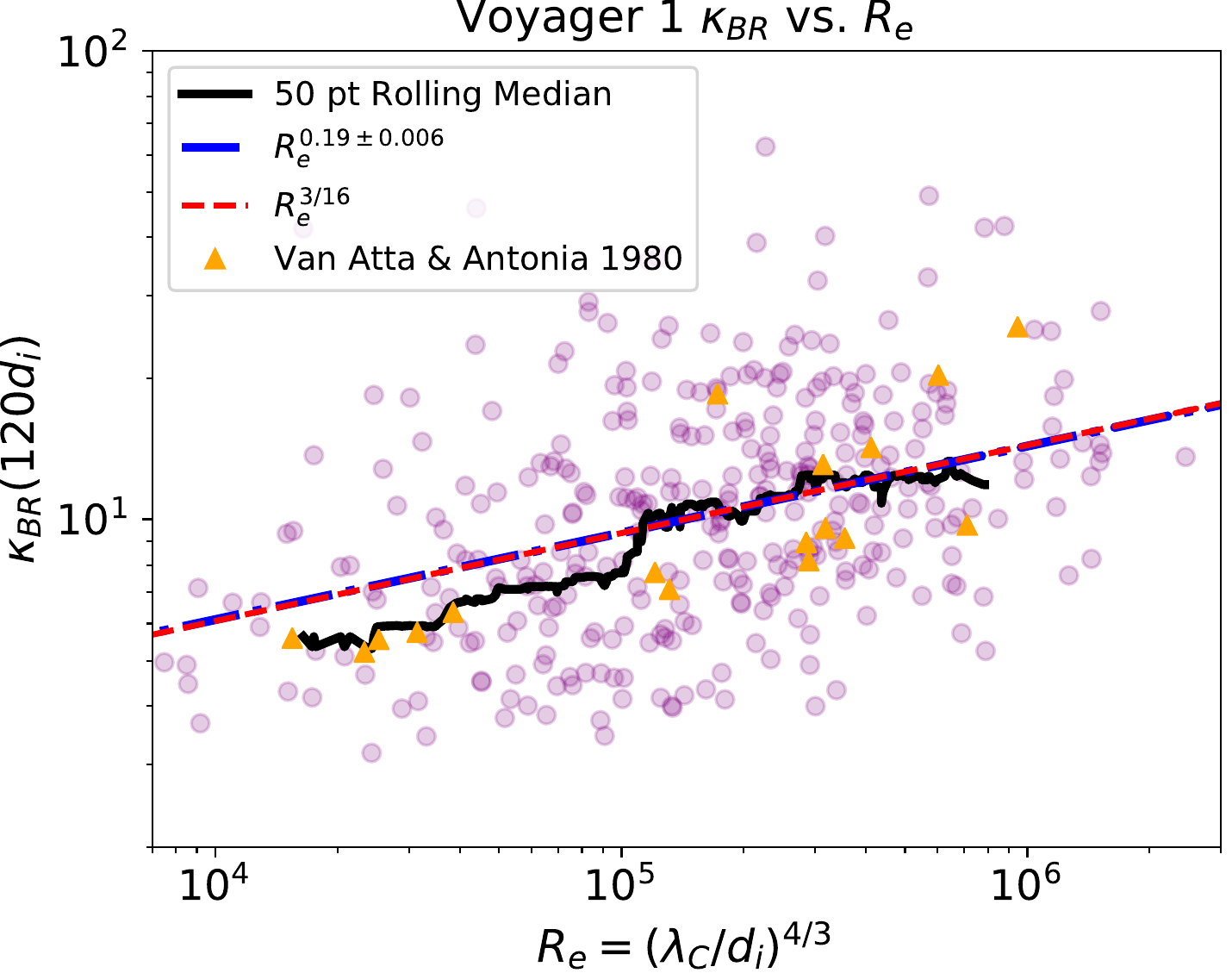}{0.3\textwidth}{(C)}
            \fig{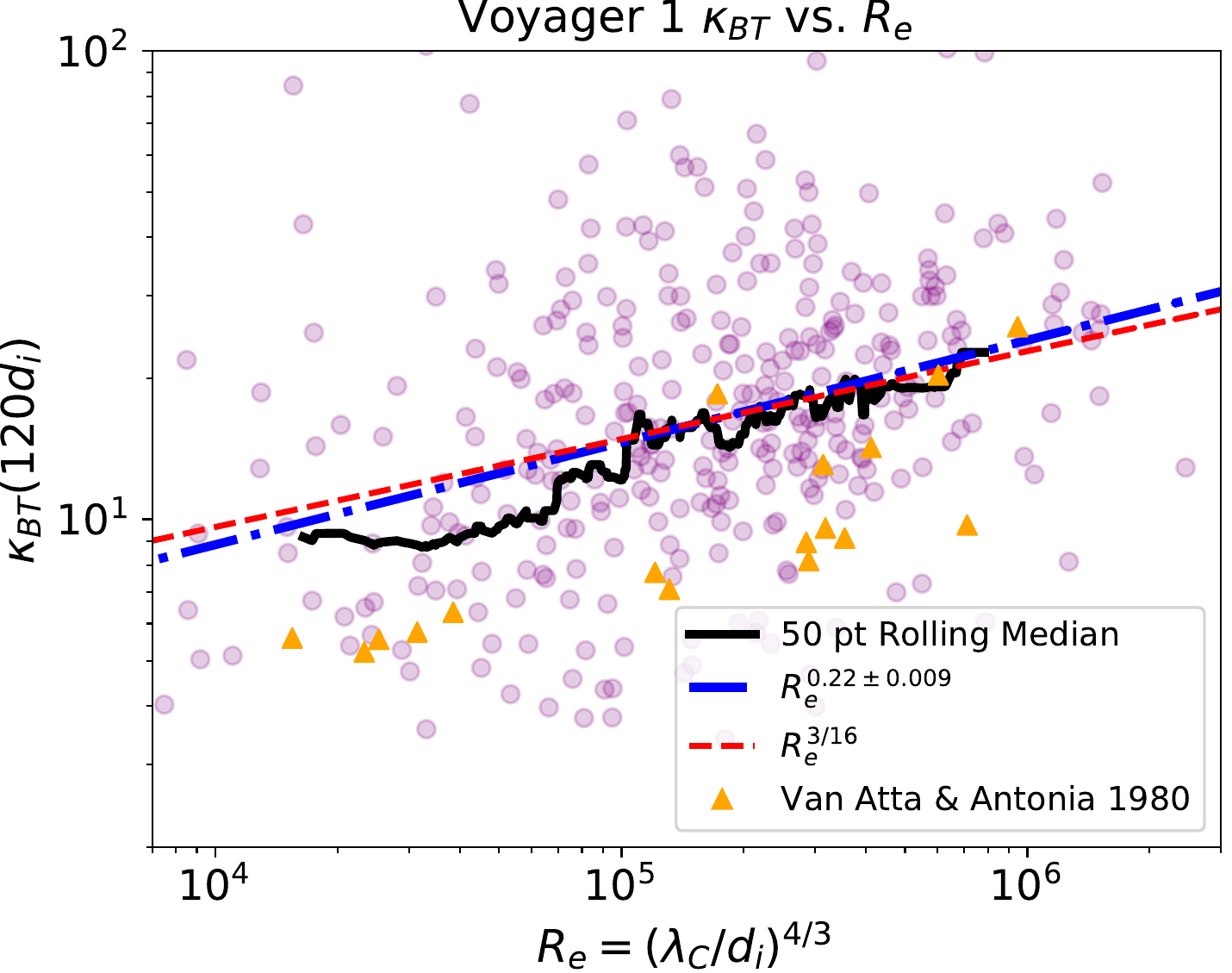}{0.3\textwidth}{(D)}
            \fig{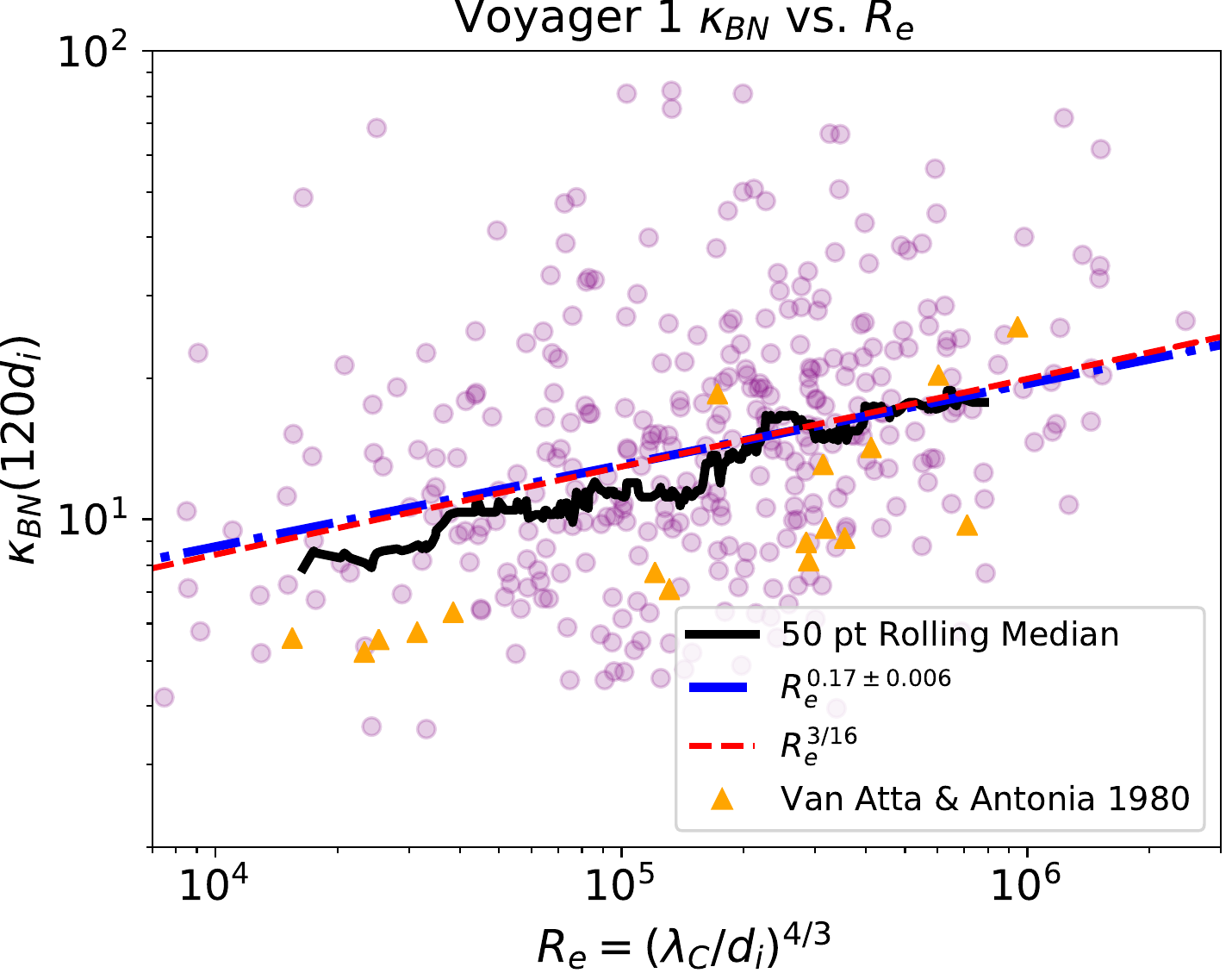}{0.3\textwidth}{(E)}
    }
    \caption{Kurtosis of the tangential and normal magnetic field fluctuating component held at $10~{\rm d_i}$ as a function of ${\rm R_e}$ for Voyager 1 (panels A and B).  Additionally, kurtosis for all three components of the fluctuating magnetic field held at $120~{\rm d_i}$ as a function of ${\rm R_e}$ (panels C, D, and E) also for Voyager 1.}
    \label{fig:vy1_KvRe}
\end{figure*}

We also provide complementary plots of the tangential and normal magnetic kurtosis as functions of effective Reynolds number, in addition to the radial magnetic kurtosis discussed in Section \ref{sec:sdkRe}.
Figures \ref{fig:hl1_KvRe} and \ref{fig:vy1_KvRe} appear to be consistent with results for $\kappa_{BR}(\ell/{\rm d_i})$ in Section \ref{sec:sdkRe}, in these specific cases of $\ell=10~{\rm d_i},120~{\rm d_i}$.
power-law fits for Figures \ref{fig:hl1_KvRe} and \ref{fig:vy1_KvRe} are computed over the range $10^5 < {\rm R_e} < 10^7$ for Helios and $10^4 < {\rm R_e} < 10^6$ for Voyager.
For more details, see Section \ref{sec:sdkRe}.

\newpage
\newpage
\newpage
\newpage

\bibliographystyle{plainnat}

%\bibliography{CuestaBib}




\end{document}